\documentclass[a4paper,11pt]{article}
\pdfoutput=1
\usepackage{jcappub}
\usepackage{tikz,xcolor,hyperref}
\usepackage{amsmath, amssymb, amsthm, graphicx, epsfig, fancyhdr,epsfig, slashed}
\usepackage[normalem]{ulem}
\usepackage{tikzsymbols}
\usepackage{natbib}
\usepackage{float}
\usepackage{bm}

\usepackage{comment}
\usepackage{graphicx}  
\usepackage{dcolumn} 
\usepackage{bm,relsize}   

\usepackage{slashed}
\usepackage{placeins}
\usepackage{adjustbox}
\usepackage[normalem]{ulem}
\usepackage{lscape}
\usepackage{multirow}

\newcommand{\GW}{\mathsmaller{\rm GW}}

\newcommand{\RH}{\mathsmaller{\rm RH}}
\newcommand{\Pl}{\mathsmaller{\rm Pl}}

\renewcommand{\eqref}[1]{(\ref{#1})}

\def\beq{\beq\begin{aligned}}
\def\eeq{\end{aligned}\eeq}

\def\beq{\begin{equation}\begin{aligned}}
\def\eeq{\end{aligned}\end{equation}}
\def\bea{\begin{eqnarray}}
\def\eea{\end{eqnarray}}

\begin{document}
%%%%%%%%%%%%%%%%%%
%%%%%%%%%%%%%%%%%%
% \begin{flushright}
%   PI/UAN-2022-718FT \\
% \end{flushright}
%%%%%%%%%%%%%%

\title{Slaying Axion-Like Particles via Gravitational Waves and Primordial Black Holes from Supercooled Phase Transition}
%%%%%%%%%%%%%%%%%%%%%%%%%%%%%%%

\author[a,b]{Angela Conaci,}
\author[a,b]{Luigi Delle Rose,}
\author[c]{P. S. Bhupal Dev,}
\author[d]{Anish Ghoshal}

\affiliation[a]{\,\,Dipartimento di Fisica, Universita’ della Calabria,\\ I-87036 Arcavacata di Rende, Cosenza, Italy}
\affiliation[b]{\,\,INFN-Cosenza, I-87036 Arcavacata di Rende, Cosenza, Italy}
\affiliation[c]{\,\,Department of Physics and McDonnell Center for the Space Sciences,\\
Washington University, St. Louis, Missouri 63130, USA}
\affiliation[d]{\,\,Institute of Theoretical Physics, Faculty of Physics, University of Warsaw,\\ ul. Pasteura 5, 02-093 Warsaw, Poland}
\emailAdd{angela.conaci@unical.it}
\emailAdd{luigi.dellerose@unical.it}
\emailAdd{bdev@physics.wustl.edu}
\emailAdd{anish.ghoshal@fuw.edu.pl}

%%%%%%%%%%%%%%%%%
\abstract{We study the formation of primordial black holes (PBHs) from density fluctuations due to supercooled phase transitions (PTs) triggered in an axion-like particle (ALP) model. We find that the mass of the PBHs is inversely correlated with the ALP decay constant $f_a$. For instance, for $f_a$ varying from ${\cal O}$(100 MeV) to ${\cal O}$($10^{12}$ GeV), the PBH mass  varies between $(10^{3} - 10^{-24}) M_{\odot}$.  We then identify the ALP parameter space where the PBH can account for the entire (or partial) dark matter fraction of the Universe, in a single (multi-component) dark matter scenario, with the ALP being the other dark matter candidate. The PBH parameter space ruled out by current  cosmological and microlensing observations can thus be directly mapped onto the ALP parameter space, thus providing new bounds on ALPs, complementary to the laboratory and astrophysical ALP constraints. Similarly, depending on the ALP couplings to other Standard Model particles, the ALP constraints on $f_a$ can be translated into a lower bound on the PBH mass scale.  Moreover, the supercooled PT leads to a potentially observable stochastic gravitational wave (GW) signal at future GW observatories, such as aLIGO, LISA and ET, that acts as another complementary probe of the ALPs, as well as of the PBH dark matter. Finally, we show that the recent NANOGrav signal of stochastic GW in the nHz frequency range can be explained in our model with $f_a\simeq (0.1~{\rm GeV}-1~{\rm TeV})$.
}
\maketitle
%%%%%%%%%%%%%%%%%%%%%%%%%%%%%%%

%%%%%%%%%%%
\section{Introduction}
\label{sec:intro}
%%%%%%%%%%%
The first direct detection of gravitational waves (GWs) by the LIGO-VIRGO collaboration~\cite{Abbott:2016blz} has opened up new avenues to explore the Universe. The known astrophysical sources of GWs can be broadly split into three categories~\cite{Maggiore:1999vm}: (i) transient signals (with a duration between a millisecond and several hours) emitted by the merger of two compact objects, like black hole or neutron star binaries, or supernova core collapse; (ii) long-duration (or steady-state) signals, e.g. from spinning neutron stars or from binary white-dwarf mergers; and (iii) stochastic background arising from the superposition of unresolved astrophysical sources. Stochastic GW background (SGWB) is also a unique probe of the early Universe, as the Universe is transparent to GWs right from the wee moments of the Big Bang, unlike other cosmic relics like photons and neutrinos. Although LIGO-VIRGO has only set an upper limit on the SGWB~\cite{LIGOScientific:2016jlg, LIGOScientific:2019vic, KAGRA:2021kbb}, the increased sensitivity of future GW detectors in a wide frequency range from nHz-kHz, such as SKA~\cite{Weltman:2018zrl}, GAIA/THEIA~\cite{Garcia-Bellido:2021zgu}, MAGIS~\cite{MAGIS-100:2021etm}, AION~\cite{Badurina:2019hst}, AEDGE~\cite{AEDGE:2019nxb}, $\mu$ARES~\cite{Sesana:2019vho}, LISA~\cite{LISA:2017pwj}, TianQin~\cite{TianQin:2015yph}, Taiji~\cite{Ruan:2018tsw}, DECIGO~\cite{Kawamura:2020pcg}, BBO~\cite{Corbin:2005ny}, ET~\cite{Punturo:2010zz},  CE~\cite{Reitze:2019iox}, as well as recent proposals for high-frequency GW searches in the MHz-GHz regime~\cite{Aggarwal:2020olq, Berlin:2021txa, Herman:2022fau, Bringmann:2023gba}, makes the future detection of SGWB an exciting real possibility. The recent evidence supporting the existence of a SGWB coming from several Pulsar Timing Arrays (PTAs), namely,  NANOGrav~\cite{NANOGrav:2023gor}, EPTA~\cite{EPTA:2023fyk},  PPTA~\cite{Reardon:2023gzh} and CPTA~\cite{Xu:2023wog}, has added fuel to the excitement and opened a floodgate of papers with various interpretations, from mundane astrophysics to exotic new physics; see e.g. Refs.~\cite{NANOGrav:2023hfp, NANOGrav:2023hvm, EPTA:2023xxk}. 

Among various cosmological mechanisms for producing  a SGWB~\cite{Roshan:2024qnv}, cosmological first-order phase transitions (FOPTs)~\cite{Caprini:2015zlo} stand out as a unique probe of beyond the Standard Model (BSM) physics, up to remarkably high scales. This is because the SM predicts only two (electroweak and QCD) phase transitions, none of which can be of first-order~\cite{Kajantie:1996mn, Bhattacharya:2014ara}. Therefore, the detection of a GW signal compatible with a FOPT would be a clear evidence of BSM physics. FOPTs develop by the formation of bubbles that expand, collide and percolate. The violent collisions between the bubble walls (and the motion of the surrounding thermal plasma) lead to the production of stochastic GWs that permeate the Universe as a relic cosmological background radiation analogous to the cosmic microwave background (CMB) radiation. 
The temperature $T$ of the plasma at the end of the FOPT is directly captured by the power spectrum of the GW signal observed today, with the frequency at the peak of the signal scaling as $f_{\rm peak}\propto T$. An FOPT at the electroweak scale of $T\sim 100$ GeV peaks around mHz~\cite{Grojean:2006bp} which is in the frequency sensitivity band of space-based GW experiments such as LISA~\cite{LISA:2017pwj}, whereas ground-based experiments such as LIGO-VIRGO~\cite{LIGOScientific:2014pky, VIRGO:2014yos} and ET~\cite{Punturo:2010zz} with a frequency band in the 100 Hz range, are capable of probing FOPTs up to $T\sim 10^8$ GeV~\cite{Dev:2016feu}, well beyond the reach of any foreseeable collider experiment.  

Interestingly, the FOPT energy scale LIGO is sensitive to roughly coincides with the lowest possible energy scale at which the global Peccei-Quinn (PQ) symmetry $U(1)_{\rm PQ}$ has to be broken in QCD axion models that solve the strong CP problem of the SM~\cite{Peccei:1977hh, Peccei:1977ur, Weinberg:1977ma,Wilczek:1977pj}. Axion can also be a viable cold dark matter (DM) candidate~\cite{Preskill:1982cy, Abbott:1982af,Dine:1982ah}. Even more generally, axion-like particles (ALPs) are well-motivated, as they naturally appear as pseudo Nambu-Goldstone bosons in many BSM extensions with a spontaneously broken global $U(1)$ symmetry, e.g. in string theory realizations~\cite{Svrcek:2006yi, Arvanitaki:2009fg, Marsh:2015xka}, in models of natural inflation~\cite{Freese:1990rb, Adams:1992bn, Daido:2017wwb}, baryogenesis~\cite{Daido:2015gqa,DeSimone:2016bok,Co:2019wyp,Jeong:2018jqe,Im:2021xoy, Foster:2022ajl} and dark energy~\cite{Jain:2004gi,Kim:2009cp,Kim:2013jka,Choi:2019jck,Brandenberger:2020gaz,Yao:2023qve}, in the relaxion mechanism for solving the hierarchy problem~\cite{Graham:2015cka}, or in unified ultraviolet-completions of the SM~\cite{Salvio:2015cja, 
Ballesteros:2016xej,Ema:2016ops,Gupta:2019ueh,Sopov:2022bog}. If the $U(1)$ symmetry breaking happens to be of first-order and strong enough, it can potentially lead to an observable SGWB at current or future GW detectors~\cite{Dev:2019njv, DelleRose:2019pgi,VonHarling:2019rgb,Ghoshal:2020vud}. In the past years, this aspect raised limited interest because the low-energy ALP phenomenology relevant for laboratory experiments does not depend on the nature of the PT. Today, however, the situation is rather different, since the opportunity of observing GW signals is concrete and offers a uniquely powerful way to {\em directly} test the high-scale PQ dynamics, in a way complementary to various laboratory and astrophysical probes of the ALP couplings to the SM particles~\cite{Choi:2020rgn} which go as the inverse of the PQ scale, with model-dependent ${\cal O}(1)$ coefficients~\cite{Kim:1979if, Shifman:1979if,Zhitnitsky:1980tq, Dine:1981rt}. Moreover, the viable parameter space for ALP masses and couplings spans many orders of magnitude, which makes it challenging to completely rule it out; therefore, new ideas and approaches to probe previously inaccessible regions are worth paying attention to. For instance, if the ALP is effectively decoupled from the SM, the only bound may come from black hole  superradiance~\cite{Cardoso:2018tly}. The GW signal from a FOPT induced by ALPs is another such probe which only depends on the $U(1)$ breaking scale but not on the ALP mass. 

In this context, the most important requirement is that the PT must be of strongly first-order so that a detectable GW signal is produced. This condition is automatically fulfilled when the theory is approximately conformal, or scale-invariant~\cite{Meissner:2006zh}. In this case, the $U(1)$ symmetry  is broken dynamically through the Coleman–Weinberg mechanism~\cite{Coleman:1973jx} and the small deviation from scale-invariance implies a suppression of the transition probability, and generically a large amount of supercooling~\cite{Witten:1980ez}. As such, bubble collisions take place in the vacuum, which increases the duration of the PT, thus enhancing the amplitude of the corresponding GW signal~\cite{DelleRose:2019pgi,VonHarling:2019rgb,Ghoshal:2020vud}.\footnote{See, for instance, Refs.~\cite{Jaeckel:2016jlh, Jinno:2016knw, Marzola:2017jzl, Iso:2017uuu, Chao:2017ilw, Baldes:2018emh, Prokopec:2018tnq, Brdar:2018num, Marzo:2018nov, Hasegawa:2019amx, Ellis:2020nnr,  Chikkaballi:2023cce, Ahriche:2023jdq} for exploitations of this mechanism in other BSM contexts. This has also been used in the context of baryogenesis via leptogenesis~\cite{Huang:2022vkf,Dasgupta:2022isg,Borah:2022cdx}, complementarity with collider searches~\cite{Dasgupta:2023zrh} and of the generation of the Planck scale~\cite{Ghoshal:2022qxk}.}

Analyzing the rich cosmological implications of such supercooled PTs for ALPs is the main objective of this work. In particular, we show that, besides an enhanced GW signal, a supercooled FOPT also leads to the formation of primordial black holes (PBHs) through the collapse of bubbles of false vacuum. In the past years, several mechanisms have been proposed regarding the formation of PBHs in the early Universe~\cite{Khlopov:2008qy}. The cosmological and astrophysical implications of PBHs can be significant~\cite{Carr:2020gox}. In particular, sub-solar mass PBHs which are formed due to the gravitational collapse of large overdensities in the primordial plasma~\cite{Carr:1974nx} may explain 100\% of the observed DM relic density of the Universe in the PBH mass range $10^{-16}M_{\odot}\lesssim M_{\rm PBH}\lesssim 10^{-10}M_{\odot}$~\cite{Carr:2016drx,Carr:2020xqk,Carr:2020gox,Green:2020jor}, where $M_\odot=2\times 10^{33}$ g is the solar mass. Even if PBHs were so light that they Hawking-evaporated~\cite{Hawking:1974rv, Hawking:1975vcx} quickly after their formation, they might have affected the DM phenomenology~\cite{Fujita:2014hha, Allahverdi:2017sks, Lennon:2017tqq, Hooper:2019gtx, Masina:2020xhk, Baldes:2020nuv, Gondolo:2020uqv, Bernal:2020bjf}. Heavier PBHs, around solar mass scale and higher, can instead contribute to the LIGO-VIRGO GW events~\cite{Bird:2016dcv, Sasaki:2016jop, Clesse:2016vqa, Hutsi:2020sol, Hall:2020daa, Franciolini:2021tla, He:2023yvl} or provide seeds for structure formation~\cite{Carr:2018rid, Liu:2022bvr, Hutsi:2022fzw}.  Therefore investigating formation mechanisms of such PBHs is interesting on its own.

During cosmic inflation, primordial overdensities may collapse into PBHs after re-entering the Hubble horizon in the post-inflationary period~\cite{Carr:1975qj}. However, for single-field inflationary scenarios, the resulting PBH abundance is exponentially sensitive to the amplitude of the curvature perturbations, and therefore it requires extreme fine-tuning~\cite{Cole:2023wyx}.\footnote{Fine-tuning may be reduced by introducing poles in the inflaton kinetic term~\cite{Ghoshal:2023pcx} or in the presence of spectator fields during inflation~\cite{Chen:2023lou}.} In this respect, other formation mechanisms such as collapsing of false vacuum bubbles, cosmic strings, domain walls or from preheating~\cite{Deng:2017uwc, Deng:2020mds, Kusenko:2020pcg, Maeso:2021xvl, Huang:2023chx, Gouttenoire:2023gbn} may reduce such fine-tuning. 
In this work, we will focus on PBH formation from supercooled FOPTs. Although this mechanism was first suggested in the pioneering works of Refs.~\cite{Hawking:1982ga,Kodama:1982sf}, it has recently received great interest due to its successful implementation in BSM scenarios with predictable results that can be tested~\cite{Lewicki:2019gmv, Ashoorioon:2020hln,Kawana:2021tde, Liu:2021svg, Jung:2021mku, Hashino:2022tcs,  Huang:2022him, Kawana:2022lba, Kawana:2022olo,Kierkla:2022odc,Kierkla:2023von}. 

The core idea is simple: FOPTs proceed via the nucleation of bubbles of the broken phase in an initial background of the symmetric phase~\cite{Coleman:1977py,Callan:1977pt,Linde:1981zj}. 
In a supercooled regime, the energy density of the Universe in the symmetric phase is dominated by the vacuum energy which acts as a cosmological constant and leads to an inflationary period. In a nucleated bubble, instead, such energy is quickly converted into radiation and the corresponding patch expands much slower.
Since bubble nucleation is a stochastic phenomenon, and since, in a supercooled regime, regions outside the nucleated bubbles expand much faster than those inside, a delayed nucleation within a causal patch can develop a large overdensity that eventually collapses, if large enough, into a PBH. The production mechanism of PBHs by such ``late-time blooming'' during strong FOPTs, was first studied in Refs.~\cite{Sato:1981bf,Kodama:1981gu,Maeda:1981gw,Sato:1981gv,Kodama:1982sf,Hsu:1990fg}, but has recently received a lot of attention~\cite{Liu:2021svg,Hashino:2021qoq,Hashino:2022tcs,He:2022amv,Kawana:2022olo,Gehrman:2023esa,Lewicki:2023ioy,Gouttenoire:2023naa,Baldes:2023rqv,Salvio:2023ynn,Banerjee:2023qya,Gouttenoire:2023pxh}, also in connection to the PTA signals observed in the form of a SGWB~\cite{NANOGrav:2023hvm,Gouttenoire:2023bqy,He:2023ado,Ellis:2023oxs}. We will study the formation of PBHs from supercooled FOPT in an explicit ALP realization, where there are only two relevant free parameters, namely, the ALP decay constant $f_a$ and a non-Abelian gauge coupling $g$. This makes the model very predictive. 

We find that the mass of the PBH scales with the square of the inverse power of $f_a$. Depending on $f_{\rm PBH}$, the fraction of DM in the form of PBHs today, 
the PBH mass is constrained over a large range, from $10^{-24} \, M_\odot$ (corresponding to $f_a \sim 10^{12}$ GeV), excluded by evaporating PBHs during Big Bang Nucleosynthesis (BBN), to $10^5 \, M_\odot$ (corresponding to $f_a \sim 10^{-2}$ GeV), excluded by CMB accretion, with an allowed window $10^{-16}M_{\odot}\lesssim M_{\rm PBH}\lesssim 10^{-10}M_{\odot}$ where $f_{\rm PBH}=1$ (100\% DM). The $f_{\rm PBH}$, on the other hand, is mainly controlled by the inverse time duration of the PT, which is related to the tunneling rate $\beta/H$ (where $H$ is the Hubble expansion rate). For values of $\beta/H \sim 5-7$, the PBH population can explain the entire DM. If the amount of supercooling is larger, namely for smaller values of $\beta/H$, the formation of PBHs can, instead, quickly overclose the Universe.

We show that the constraint from the overproduction of PBHs, as well as the bounds on their mass from BBN, CMB and microlensing~\cite{PBHbounds}, can strengthen the interplay between GW searches and laboratory, astrophysical and cosmological constraints on the ALP decay constant $f_a$~\cite{AxionLimits}. Through this three-pronged complementarity, we are able to severely cut down the allowed ALP parameter space. Specifically, the GW sensitivity reach of future interferometers and the PBH parameter space ruled out by the aforementioned constraints can be directly mapped onto the ALP parameter space, which provides new bounds on ALP couplings. Conversely, depending on the strength of ALP interactions to SM particles, the bound on the ALP decay constant can be translated into a lower bound on the PBH mass scale. As a side note, we also show that the recent NANOGrav signal of stochastic GW in the nHz frequency range can be explained in our model with $f_a\simeq (0.1~{\rm GeV}-1~{\rm TeV})$ while satisfying the PBH constraints.

The rest of this article is organized as follows: in Section~\ref{sec:model} we discuss supercooled PT in the ALP scenario, provide an explicit realization and identify the parameter space responsible for supercooling due to radiative symmetry breaking. In Section~\ref{sec:SupercoolGW}, we discuss the predictions for the GW signal. In Section~\ref{sec:SupercoolPBH}, we discuss the PBH formation. In Section~\ref{sec:results} we present our numerical results. In Section~\ref{sec:complementarity} we show the three-pronged complementarity of ALP phenomenology following PBH, GW and laboratory/astrophysical searches. Our conclusions are given in Section~\ref{sec:conclusions}. Some supplementary plots for $f_{\rm PBH}\ll 1$ are given in Appendix~\ref{ref:supp}.

%%%%%%%%%%%
\section{The supercooled ALP phase transition}
\label{sec:model}
%%%%%%%%%%%

We consider a scenario with spontaneous radiative breaking of a global $U(1)$ symmetry~\cite{Gildener:1976ih}. To realize this scenario, we consider a collection of massless scalars, some of which charged under the $U(1)$ symmetry, described by the tree level potential
\bea
V(\phi) = \frac{\lambda_{ijkl}}{4} \phi_i \phi_j \phi_k \phi_l \,.
\eea 
A shown in Ref.~\cite{Gildener:1976ih}, the renormalization group equations of the quartic couplings generically imply that a linear combination of those vanishes at some energy scale $\Lambda$. 
At that scale, one can identify a flat direction in the potential parameterized by a unit vector $\vec{n}$. In the field coordinates $\vec{\phi} = \vec{n} \sigma$, the tree-level potential vanishes identically in the direction of $\sigma$. As such, the dynamics of the system along $\sigma$ is fully controlled by the one-loop radiative corrections and the effective potential can be recast in the form
\bea
\label{eq:effpot0}
V_\textrm{eff}(\sigma) = \frac{\beta_{\lambda_\textrm{eff}}}{4} \sigma^4 \left( \log \frac{\sigma}{f}  - \frac{1}{4}\right) \, ,
\eea
where $\beta_{\lambda_\textrm{eff}}$ is the $\beta$ function of the effective quartic coupling $\lambda_\textrm{eff}(\mu) = \lambda_{ijkl}(\mu) n_i n_j n_k n_l$ satisfying the condition $\lambda_\textrm{eff}(\Lambda) = 0$. For positive $\beta_{\lambda_\textrm{eff}}$, the effective potential has a minimum at $f$ and the $\sigma$ field acquires a mass $m_\sigma^2 = \beta_{\lambda_\textrm{eff}} f^2$.
\subsection{Finite-temperature effects} \label{sec:2.1}
%%%%%%%%%%
Around the origin, $\sigma \simeq 0$, the potential $V_\textrm{eff}$ is flat (since the second-order derivative is vanishing) and the finite temperatures effects become extremely relevant~\cite{Witten:1980ez}, inducing a quadratic positive curvature for any $T > 0$, thus effectively turning the local maximum at $\sigma=0$ into a local metastable minimum. These thermal effects are described by the finite-temperature potential
\bea
V_T(\sigma, T) = \frac{T^4}{2 \pi^2} \sum_{\rm b} J_B\left( \frac{m_{\rm b}^2(\sigma)}{T^2} \right) + \frac{T^4}{2 \pi^2} \sum_{\rm f} J_F\left( \frac{m_{\rm f}^2(\sigma)}{T^2} \right) \,,
\eea
where $m_{\rm b,f}(\sigma)$ are the field-dependent masses of all the relevant bosonic and fermionic fields, and $J_{B,F}$ are the corresponding thermal functions given by 
\begin{align}
    J_{B/F}(y^2) = \int_0^\infty dt~t^2 \log{\left[1\mp \exp{\left(-\sqrt{t^2+y^2}\right)}\right]} \, .
    \label{eq:JBF}
\end{align}
Near the origin, due to the vanishing of the second-order derivative, the high-temperature expansion ($y^2\ll 1$) is always well-justified. In this limit, the free-energy of the system along the $\sigma$ direction can be written as
\footnote{Bosonic degrees of freedom also provide a cubic term in the $m^2/T^2$ expansion which has a mild numerical impact on the computation of the temperature-dependent bounce action in supercooled PTs~\cite{DelleRose:2019pgi,Salvio:2023qgb}.}
\bea
F(\sigma, T) \simeq - \frac{\pi^2}{90} g_* T^4 +  a \frac{T^2}{24} \sigma^2  + \frac{\beta_{\lambda_\textrm{eff}}}{4} \sigma^4 \left( \log \frac{\sigma}{f}  - \frac{1}{4}\right) + V_0 
\eea
where $g_* = N_{\rm b} + \frac{7}{8} N_{\rm f}$ represents the effective number of relativistic degrees of freedom in the meta-stable phase, the coefficient $a$ is the defined as $a \sigma^2 \equiv \left[ \sum_{\rm b} m_{\rm b}^2(\sigma)  +  \frac{1}{2} \sum_{\rm f} m_{\rm f}^2(\sigma) \right]$ and $V_0$, given by $V_0 = - V_\textrm{eff}(\sigma = f) = \beta_{\lambda_\textrm{eff}} f^4/16$, is chosen to remove the cosmological constant in the global minimum.

Away from the origin, the high-temperature expansion used above is valid provided that $ \sigma \lesssim T/{\hat g}$ with $\hat g$ being the typical coupling constant in the field-dependent mass $m(\sigma) \sim \hat g \sigma$. \\
During a supercooled PT, the free-energy above can be further simplified by noting that
\bea
\log \frac{\sigma}{f} = \log \frac{\hat g \sigma}{T} + \log \frac{T}{\hat g f}  \simeq \log \frac{T}{\hat g f} \equiv  \log \frac{T}{M} \,,
\eea
with $M$ being the typical heavy mass scale at the minimum. Indeed, while during supercooling $T \ll M$, the bounce action will get much of its contribution from field values around the barrier, for which $\hat g \sigma/T \sim 1$. In this case the free-energy can be recast in a simple polynomial form with a positive quadratic and a negative quartic term\footnote{We can check a posteriori the validity of the approximation by noting that the potential in Eq.~(\ref{eq:approxpot}) provides a barrier size $\sigma_b = \sqrt{2 m^2(T)/\lambda(T)} = T \sqrt{a/(6 \beta_{\lambda_\textrm{eff}} \log(M/T))}  < T/ \hat g $ due to the large logarithm.}
\bea
\label{eq:approxpot}
F(\sigma, T) \simeq \frac{m^2(T)}{2} \sigma^2 - \frac{\lambda(T)}{4} \sigma^4
\eea
with $m^2(T) = a T^2/12$,  $\lambda(T) = \beta_{\lambda_{\textrm{eff}}} \log (M/T)$ and the field-independent terms omitted here are  irrelevant for the computation of the transition rate.

The PT of the model in Eq.~(\ref{eq:approxpot}) is controlled by the 3-D bounce action $S_3/T \approx 18.897 \, m(T)/(\lambda(T) T) \equiv A_3/\log(M/T)$, which depends only logarithmically on the temperature as a result of the approximate scale invariance of the model. The corresponding tunneling rate is given by
\bea
\Gamma \simeq T^4 \left( \frac{S_3}{2\pi T} \right)^\frac{3}{2} \exp(-S_3/T) \,,
\eea
and the slow (logarithmic) temperature dependence in the bounce action generically implies a phase of supercooling. 
Beside tunneling at finite temperature, nucleation of true vacuum bubbles can also be
driven by 4-D bounces. If the $O(4)$ bounce action $S_4$ is smaller than $S_3/T$, quantum effects can
lead to a faster nucleation rate but we find that the thermal nucleation rate always dominates in the relevant regions of our parameter space.

The nucleation temperature is defined by $\Gamma/H^4 = 1$, with the Hubble rate given by
\bea
H^2 = \frac{1}{3M_\textrm{Pl}^2} \left[ \frac{\pi^2}{30} g_* T^4  +  V_0 \right] \simeq \frac{V_0}{3M_\textrm{Pl}^2} \equiv H_I^2 \,,
\label{eq:2p9}
\eea
where $M_{\rm Pl}=2.4\times 10^{18}$ GeV is the reduced Planck mass, and the right-hand side of Eq.~\eqref{eq:2p9} holds during the supercooling phase, under which the Universe becomes vacuum-dominated and experiences an inflationary period. Using the equation above, the nucleation temperature is found to be
\bea
T_n \simeq \sqrt{M H_I} \exp \left( \frac{1}{2} \sqrt{-A_3 + \log^2(M/H_I)} \right) \,,
\eea
with a lower bound given by $T_n^\textrm{min} = \sqrt{M H_I}\simeq 0.1f(f/M_{\rm Pl})^{1/2}$. 

The strength $\alpha$ of the PT can be characterized by the free energy difference between the metastable and true vacua, normalised to the radiation density. During supercooling, the vacuum energy dominates and we have
\bea
\alpha = \frac{1}{\rho_\textrm{rad}} \left( 1 - T \frac{\partial}{\partial T} \right) \bigg[ F(0, T) - F(\sigma_n, T)  \bigg]\bigg|_{T_n} \simeq \frac{V_0}{\rho_\textrm{rad}(T_n)} = \left( \frac{T_\textrm{inf}}{T_n} \right)^4 \,,
\eea
with $T_\textrm{inf} = (30 V_0/(g_* \pi^2))^{1/4}$ being the temperature at which the Universe becomes vacuum-dominated.\footnote{We assume that $g_*$ does not change between $T_\textrm{inf}$ and $T_n$.}

Finally, the logarithmic derivative of the tunneling rate is found to be
\bea
\frac{\beta}{H} = - \frac{d \log \Gamma}{d \log T} \bigg|_{T=T_n} \simeq - 4 + T \frac{d (S_3/T)}{d T}\bigg|_{T=T_n} = - 4  + \frac{A_3}{\log^2(M/T_n)} \,,
\eea
which clearly shows that, in the supercooling regime, $\beta/H$ can become of order $O(1)$ thus maximizing the power spectrum of GW as well as the production of PBHs.

After the completion of the PT the Universe is reheated at a temperature 
\bea
T_{\rm RH} = T_\textrm{inf} \min\left(1, \frac{\Gamma}{H_I} \right)^{1/2} \,,
\eea
where $\Gamma$ is the relevant decay rate into the SM plasma. For the sake of simplicity, in the following we will assume sufficiently fast reheating, namely $T_{\rm RH} \simeq T_\textrm{inf}$.

Within the approximations discussed above, the dynamics is controlled by the three parameters $f$, $\beta_{\lambda_\textrm{eff}}$ and $a$ (see also Ref.~\cite{Salvio:2023qgb}) which can be completely specified once an explicit realization of the model is provided. 

\subsection{An explicit ALP realization}
%%%%%%%%%%%
As a prototype for an elementary realization of the radiative breaking of the global $U(1)$ symmetry, let us consider a pair of complex scalar fields $\phi_1$ and $\phi_2$, neutral under the SM gauge group, a pair of vector-like fermions $\psi, \psi^c$, and an $SU(2)$ gauge field $A^\mu$, with gauge coupling $g$. Under the new $SU(2)$ symmetry only $\phi_1$ is charged, while being neutral under the global $U(1)$. The $\phi_2$ transforms, instead, under the global $U(1)$ and its phase corresponds to the ALP. The $SU(2)$ gauge sector controls the running of the effective quartic coupling of the scalar potential. The simpler choice of an abelian gauge group would have worked equally well. We opted for the non-abelian scenario to avoid possible issues associated to the Landau pole and further complications related to the presence of cosmic strings. The latter might actually enrich the phenomenology of the model and their interplay will be addressed in a separate work.

The Lagrangian of the model is given by
\bea
\label{eq:elemLag}
\mathcal L = - \frac{F^2}{4 g^2} + |D_\mu \phi_1|^2 + |\partial_\mu \phi_2|^2 + (y \phi_2 \psi \psi^c + \textrm{h.c.} ) - \lambda_1 |\phi_1|^4 - \lambda_2 |\phi_2|^4  -  \lambda_{12} |\phi_1|^2 |\phi_2|^2 \, ,
\eea
where $F^{\mu\nu}$ and $D^\mu$ are the field strength and the covariant derivative of the $SU(2)$ gauge field, respectively. 
For the sake of simplicity, in order to disentangle the radiative global $U(1)$ breaking from the electroweak symmetry breaking, we neglect the tree-level portal couplings to the SM Higgs doublet. This class of models has been studied in Refs.~\cite{DelleRose:2019pgi,VonHarling:2019rgb} in the context of supercooled QCD-axion models and similar realizations were also addressed in the context of electroweak PTs  in Refs.~\cite{Hambye:2013dgv,Iso:2017uuu,Hambye:2018qjv}. Another interesting scenario is characterized by a composite ALP arising from the spontaneous breaking a global symmetry of a strongly coupled dynamics. In this context, a supercooled PT can be realized within a strongly coupled conformal sector with the spontaneous breaking of scale invariance~\cite{DelleRose:2019pgi,VonHarling:2019rgb}. For the sake of definiteness, in this paper we focus on an elementary ALP scenario.

In the potential of Eq.~(\ref{eq:elemLag}), the flat direction of the tree-level potential is found at a scale $\Lambda$ at which $\lambda_{12} = - 2 \sqrt{\lambda_1 \lambda_2}$ and, in the manifold of the scalar fields, is parameterized by 
\bea
\label{eq:defmodel}
\phi_i = (\cos \theta, \sin \theta) \frac{\sigma}{\sqrt{2}} \,, \qquad \textrm{with}  \qquad \tan^2 \theta = \sqrt{ \frac{\lambda_1}{\lambda_2} } \,.
\eea
Along the flat direction $\sigma$, the field-dependent masses of the gauge field $A^\mu$, the fermion $\psi$ and the scalar field $\tau$ in the radial direction orthogonal to $\sigma$ are respectively given by
\bea
M_A = g \, \cos \theta \, \sigma \,, \qquad M_\psi = \frac{y}{\sqrt{2}} \sin \theta \, \sigma \,, \qquad M_\tau = (4 \lambda_1 \lambda_2)^{1/4} \sigma \,.
\eea
In the previous list we did not count the phase of $\phi_2$, the ALP, being exactly massless at this level, and the $\sigma$, the mass of which is loop-suppressed. With the rotation angle $\theta$ introduced in Eq.~(\ref{eq:defmodel}), the axion decay constant $f_a$ is given by $f_a = f \sin \theta$.
The effective potential in the direction of $\sigma$ takes the form of $V_\textrm{eff}$ in Eq.~(\ref{eq:effpot0}) with
\bea
\beta_{\lambda_\textrm{eff}} = \frac{\partial \lambda_\textrm{eff}}{\partial \log \mu} =  \frac{1}{16 \pi^2} \left[ 6 N g^4 \cos^4 \theta - 2 y^4 \sin^4 \theta  + 8 \lambda_1 \lambda_2 \right] \,,
\eea
where $N=3$ is the number of gauge bosons of $SU(2)$. Analogously, we can compute the thermal corrections and, in particular, we can extract the coefficient of the quadratic terms in high-temperature regime:
\bea
a = 3 N g^2 \cos^2 \theta + 3 y^2 \sin^2 \theta + 2 \sqrt{\lambda_1 \lambda_2} \,.
\eea
In the following, we will address the scenario in which the gauge coupling dominates, namely, $\lambda_1 \simeq \lambda_2 \ll g^2$ and $y \simeq 0$.

\begin{figure}[t!]
\centering
\includegraphics[width=0.45\textwidth]{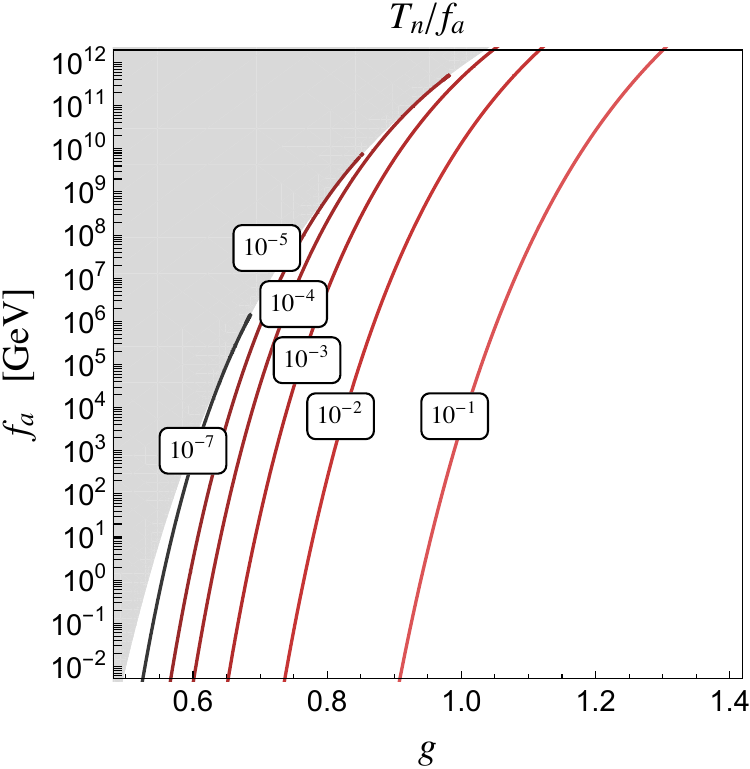}
\hspace{0.2cm}
\includegraphics[width=0.45\textwidth]{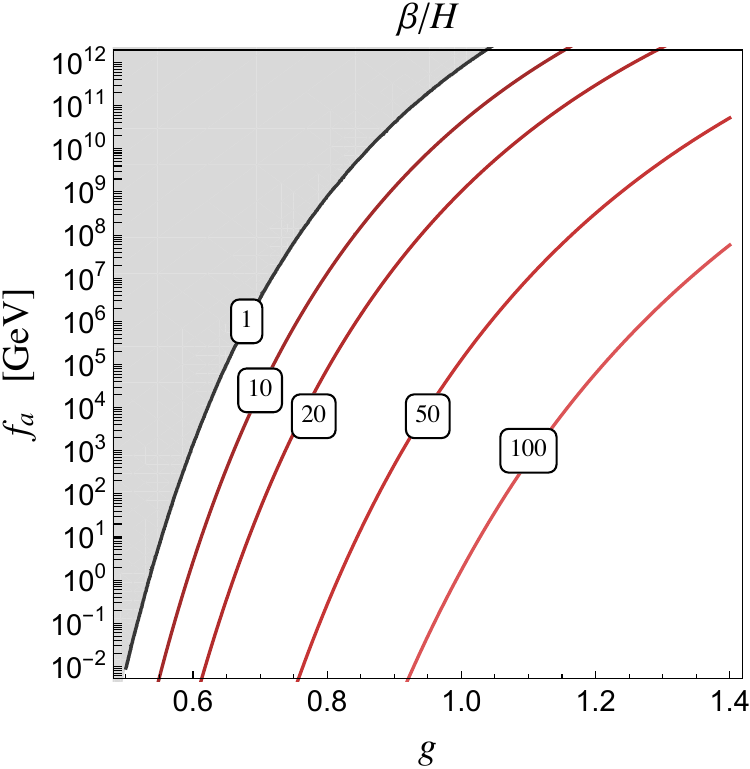}
\caption{Contours showing the nucleation temperature $T_n$ (in units of $f_a$) and the $\beta/H$ parameter in the plane of the model parameters $f_a$ and $g$. In the gray region nucleation never occurs.} 
\label{fig:bH}
\end{figure}

In Fig.~\ref{fig:bH} we show the contours of the nucleation temperature (left panel) and the $\beta/H$ parameter (right panel) in the parameter space of the model discussed here, entirely defined by the gauge coupling $g$ and the axion decay constant $f_a$. In the present work, we focus on the supercooling regime, approximately determined by $T_n/f_a \lesssim 10^{-2}$. In this case, the dependence of $T_n$, $\beta/H$ and $\alpha$ is well described by the analytic approximations discussed in Section~\ref{sec:2.1}. In the supercooling regime, $\alpha$ is always very large, $\alpha \gg 1$, and basically drops out from the expressions of both the GW spectrum and the PBH abundance that will be discussed below. For this reason, we do not show the dependence of $\alpha$.

In the gray region of Fig.~\ref{fig:bH}, the system remains stuck in the inflationary phase. Nevertheless, the de Sitter fluctuations in the false vacuum may actually push the field into the true ground state, thus effectively completing the transition. This happens when the Hubble constant becomes larger than the height of the barrier. To describe this process, once should take into account the de Sitter curvature~\cite{Kearney:2015vba,Joti:2017fwe} in the computation of the tunneling rate. We do not address this case in our work, and will therefore stick to the unshaded region of parameter space in Fig.~\ref{fig:bH}. 

%%%%%%%%%%%
\section{GW signals from supercooled ALP phase transition}
\label{sec:SupercoolGW}
%%%%%%%%%%%
The spectra for GW signal is as usual given by
	\begin{equation}
	h^{2}\Omega_{\GW}(f) \equiv \frac{h^{2}}{ \rho_{c}} \frac{ d \rho_{\GW}}{ d \log f }
	\end{equation}
where $\rho_{\GW}$ is the energy density in GWs, $\rho_{c}$ is the critical density of the Universe, and $h$ is the dimensionless Hubble parameter\footnote{The generation of Gravitational Waves occur during the bubble collision, sourced by the non-zero anisotropic stress component in their space-space part $\Pi_{i j}$, see Refs. \cite{Caprini:2015zlo,Caprini:2018mtu,Athron:2023xlk} for a review.}.
After the GWs are produced in the early Universe they are red-shifted and we observe them today with an amplitude~\cite{Kosowsky:1992rz}
	\begin{subequations}
	\begin{align}
	h^{2}\Omega_{\GW} & = h^{2}\Omega_{\rm rad}(T_0) \left( \frac{ g_{\ast s}(T_0) }{ g_{\ast s}(T_{\RH}) } \right)^{1/3} \Omega_{\mathrm{GW} \ast}  \\
	                  & = \frac{ g_{\ast}(T_0) \pi^{2} T_0^{4}}{30  } \frac{1}{ 3 H_{100}^{2} M_{\Pl}^2}  \left( \frac{ g_{\ast s}(T_0) }{ g_{\ast s}(T_{\RH}) } \right)^{1/3} \,  \Omega_{\mathrm{GW} \ast} \\
		          & =  6.56 \times 10^{-5} \,  \left(g_{\ast s}(T_{\RH})\right)^{-1/3} \,  \Omega_{\mathrm{GW} \ast},
	\end{align}
	\end{subequations}
 where we used $\Omega_{\rm rad}(T_{\RH})  \equiv \rho_{\rm rad}/\rho_{c}  = 1$ just after the FOPT and
 $\Omega_{\mathrm{GW} \ast}$ is the GW signal just after the PT, $T_{0} \simeq 0.23$ meV is the CMB temperature today, and $H_{100} \equiv 100$~km/s/Mpc. Similarly, the frequency is also red-shifted as
	\begin{equation}
	f_{0} = \left( \frac{ g_{\ast s}(T_0) }{ g_{\ast s}(T_{\RH}) } \right)^{1/3} \frac{ T_{0} }{ T_{\RH} } \, f_{\ast} \,,
	\end{equation}
with $f_{\ast}$  denoting the frequency of the GW spectrum at the PT. The reheating that occurs at the end of the FOPT is characterized by the temperature, via the first Friedmann equation:
	\begin{equation}
	T_{\RH} = \left( \frac{ 90M_{\Pl}^{2}H(T_{\RH})^2 }{ g_{\ast}(T_{\RH})\pi^2 } \right)^{1/4} \,.
	\end{equation}
Notice also that, for supercooled PTs, we will deal with highly relativistic bubble walls. 

For the estimates of the spectral shape of GWs from a FOPT we use the results from the hybrid simulations obtained in Ref.~\cite{Lewicki:2020azd} in which the anisotropic stress induced in a bubble collision was first determined in a (1+1)-dimensional simulation. The interesting aspect of the analysis is that there are notable differences in the GW spectra between non-gauged and gauged scenarios (see Ref.~\cite{Lewicki:2020jiv} for the non-gauged case and Ref.~\cite{Lewicki:2020azd} for the gauged case  of U(1) symmetry breaking\footnote{Simulation with non-Abelian gauge theory has not been done; however, we do not expect much to change as the GW spectrum varies very mildly and does not impact our analysis too much.}). Since we have a gauged theory, the GW spectrum is characterized by 
	\begin{equation}
	h^{2}\Omega_{\rm hyb}(f) = 5.10 \times 10^{-9} \left( \frac{ 100 }{  g_{\ast }(T_{\RH})} \right)^{1/3} \left( \frac{ 10 }{ \beta/H } \right)^{2}  S_{\rm hyb}(f),
	\end{equation} 
with the shape,
	\begin{equation}
	S_{\rm hyb}(f) = \frac{ 695 }{ \left[ 2.41 \left( \frac{f}{\tilde{f}_{\rm hyb}} \right)^{-0.557} + 2.34 \left(  \frac{f}{\tilde{f}_{\rm hyb}} \right)^{0.574} \right]^{4.20} } \, ,
	\end{equation}
and peak frequency
	\begin{equation}
	\tilde{f}_{\rm hyb} =  1.1\times 10^{-4} \; \mathrm{Hz} \, \left( \frac{ g_{\ast} }{ 100 } \right)^{1/6} \left(  \frac{ \beta/H }{ 10 } \right)   \, \left( \frac{ T_{\RH} }{ 500 \; \mathrm{GeV} } \right).
	\end{equation}
The expressions above are only valid for the FOPT we are interested in, that is supercooled PT. This basically means we are considering the limit $\alpha \gg 1$, and ultra-relativistic bubble-wall velocity $v_{w} \simeq 1$. We also enforce the correct scaling $h^{2}\Omega_{\rm GW} \sim f^3$ for super-horizon modes which correspond to the frequencies at IR tail of the spectrum below~\cite{Durrer:2003ja,Caprini:2009fx,Barenboim:2016mjm,Cai:2019cdl,Hook:2020phx}
\begin{equation}
    f_{\rm horizon} = 0.013 \, {\rm mHz} \left( \frac{ g_{\ast} }{ 100 } \right)^{1/6} \left( \frac{ T_{\RH} }{ 500 \; \mathrm{GeV} }  \right) \,. 
\end{equation}
Other estimates of the GW spectral shapes are available in the literature, e.g. the one obtained using $(3+1)$-dimensional lattice simulations of thick wall bubble collisions~\cite{Cutting:2020nla} and the semi-analytical estimates developed in Refs.~\cite{Jinno:2017fby,Konstandin:2017sat}. We checked that all these estimates provide similar results after requiring the correct scaling $h^{2}\Omega_{\rm GW} \sim f^3$ for super-horizon modes.

%%%%%%%%%%%
\section{PBH formatiom from supercooled ALP phase transition}
\label{sec:SupercoolPBH}
%%%%%%%%%%%

%%%%%%%%%%%%%%%%%%%%%%%%%%%%%%%%%%%%%%%%%%%%%%%%%%%%%%

PBHs can be formed due to the collapse of overdense regions during a strong FOPT~\cite{Liu:2021svg,Kawana:2022olo,Gouttenoire:2023naa,Lewicki:2023ioy}. During a supercooled PT, when the plasma temperature reaches $T_{\rm inf}$, the Universe enters a vacuum-dominated era and a phase of inflation takes place. 
This period lasts until $T_n$, when bubble nucleation becomes efficient. 
Then, regions of the Universe undergo a transition to the true ground state and get reheated at $T_{\rm RH} \simeq T_{\rm inf}$. 
Since nucleation is a stochastic process, some Hubble patches may remain in the false vacuum and inflate more.

When bubble nucleation begins, and afterwords during bubble growth, the vacuum energy is converted into energy stored in the bubble wall, energy dumped into the plasma and sound waves from bubble collisions, all of them redshifting away as radiation, decreasing as $\propto a^{-4}(t)$.
On the other hand, casual patches in which nucleation is delayed still inflate due to the constant, and dominating, vacuum energy contribution. Eventually, they generate an overdensity with respect to the surrounding average background.

The ratio $\rho_{\rm rad}^{\rm late}/\rho_{\rm rad}^{\rm bkg}$ between the energy densities of the late-blooming patch and of the background keeps increasing until it reaches a maximum value, approximately determined by the percolation temperature of the former.
If the overdensity of the late-nucleated patch, quantified by the density contrast $\delta = \rho_{\rm rad}^{\rm late}/\rho_{\rm rad}^{\rm bkg} - 1$, is larger than a critical threshold, usually $\delta_c \simeq 0.45$, the patch can collapse into a PBH~\cite{Musco:2004ak} (see Refs.~\cite{Kawana:2022olo,Liu:2021svg,Gouttenoire:2023naa,Lewicki:2023ioy} for more details\footnote{There are minor differences in estimation of nucleation time between the studies in Ref.~\cite{Kawana:2022olo, Liu:2021svg, Gouttenoire:2023naa, Lewicki:2023ioy} based on contribution from past light cones which may lead to slightly varied abundances of PBHs.}).
We do not go into the details of such computations and instead follow Ref.~\cite{Gouttenoire:2023naa} for our estimates\footnote{We follow the analysis of this section, satisfying Hoop's conjecture for PBH formation as discussed in Refs.~\cite{Lewicki:2023ioy,Lewicki:2024ghw}.}.

The probability that a Hubble patch collapses into a PBH can be approximated, in the limit of $\alpha \gtrsim 10^2$, by the analytic formula
\bea
P_{\rm coll} \simeq \exp \left[ - a \left(\frac{\beta}{H_n} \right)^b (1+ \delta_c)^{c \frac{\beta}{H_n}} \right] \,,
\eea
where the coefficients $a \simeq 0.5646$, $b \simeq 1.266$, $c \simeq 0.6639$ are fitted from a numerical computation. The collapsing probability does not depend on the scale of the PT, but only on the $\beta/H_n$ parameter and on the critical threshold $\delta_c$.

The mass of the PBH is given by the energy contained within the sound horizon $c_s H^{-1}$ at the time of the collapse $t_{\rm coll}$~\cite{Escriva:2021pmf}:
\bea
M_{\rm PBH} = k \frac{4\pi}{3} c_s^3 H^{-3}  \rho_{\rm rad}^{\rm late} \bigg|_{t_{\rm coll}} \simeq 3.7 \times 10^{-8} M_{\odot} \left( \frac{106.75}{g_*(T_{\rm RH})} \right)^{1/2}  \left( \frac{500 \, {\rm GeV}}{T_{\rm RH}} \right)^2 \,.
\eea
Here $k$ ($\le 1$) is a numerical factor which depends on the details of the gravitational collapse involved in such a process and $M_{\odot}$ is the solar mass.

\begin{figure}[t!]
\centering
\includegraphics[width=0.45\textwidth]{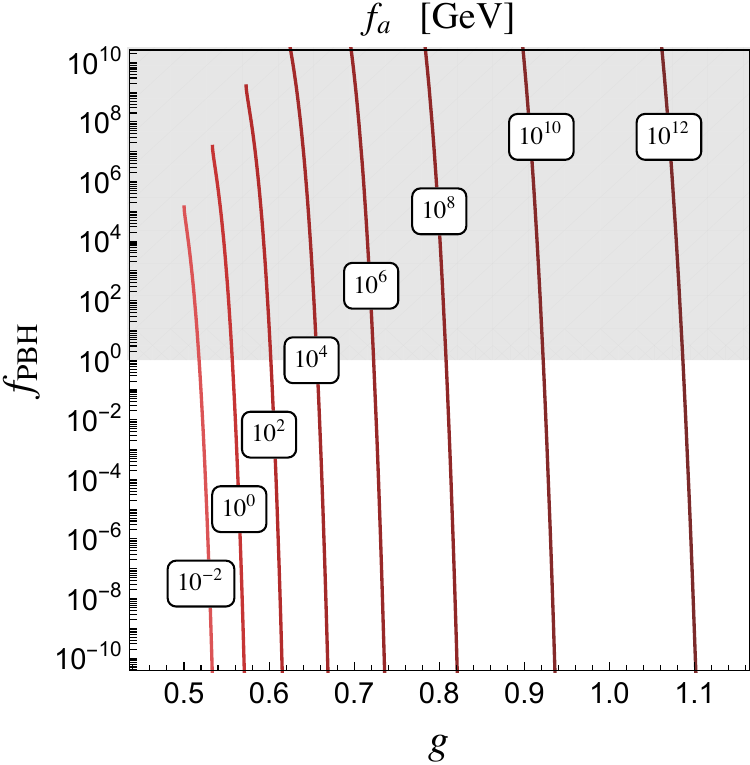}
\hspace{0.2cm}
\includegraphics[width=0.45\textwidth]{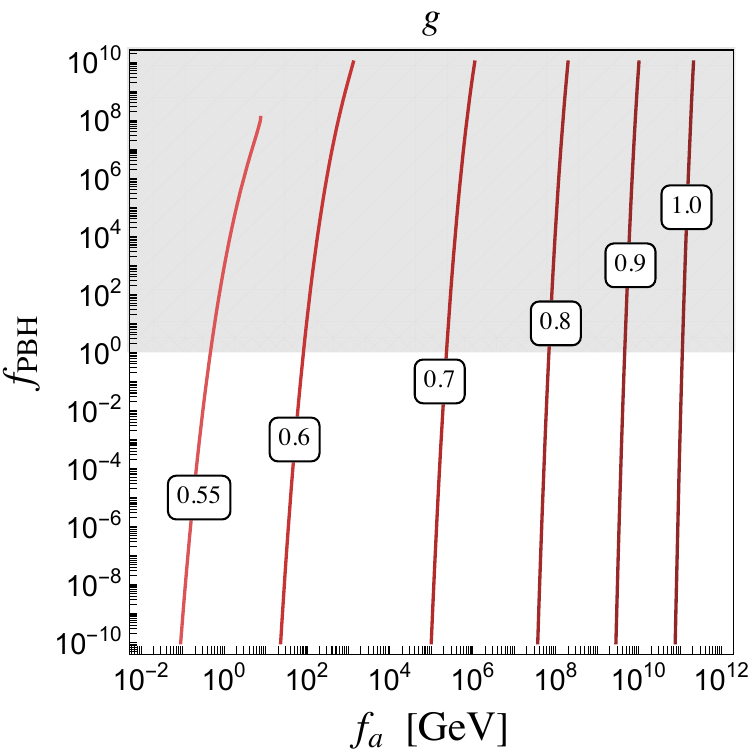}
\caption{The fraction of DM in the form of PBH, $f_{\rm PBH}$, as functions of the two parameters of the model, $g$ (left panel) and $f_a$ (right panel), with constant contours of the other parameter shown. The gray region corresponds to PBH overclosure. } 
\label{fig:fPBH}
\end{figure}

Finally, the fraction of DM in the form of PBHs today is found to be
\bea
f_{\rm PBH} \equiv \frac{\rho_{\rm PBH, 0}}{\rho_{\rm DM,0}} = P_{\rm coll} \frac{M_{\rm PBH} \, \mathcal N_{\rm patches}}{\frac{4 \pi}{3} H_0^{-3}\rho_{\rm DM,0}}
\simeq \left( \frac{P_{\rm coll}}{6.2 \times 10^{-12}} \right) \left( \frac{T_{\rm RH}}{500 \, {\rm GeV}} \right) 
\eea
where $\rho_{\rm DM,0} \simeq 0.26 \times 3 M_{\rm pl}^2 H_0^2$ is the current DM energy density and ${\mathcal N}_{\rm patches}$ represents the number of Hubble patches, when the temperature was $T_{\rm RH}$, in our past light-cone, namely,
\bea
\mathcal N_{\rm patches} = \left( \frac{a_{\rm RH} H_{\rm RH}}{a_0 H_0 } \right)^3 \simeq 5.3 \times 10^{40} \left( \frac{g_*(T_{\rm RH})}{100} \right)^{1/2} \left( \frac{T_{\rm RH}}{500\,{\rm GeV}} \right)^3.
\eea
The PBH abundance is shown in Fig.~\ref{fig:fPBH} as functions of the two parameters of the model, $f_a$ and $g$.

The overdense regions generated from the late decay of the false vacuum in a supercooled FOPT also produce curvature perturbations~\cite{Liu:2022lvz}. These can be described assuming a Gaussian distribution. If the distribution deviates from monochromaticity, the collapsing region can have a non-zero spin, although typically small for a FOPT.
Despite that, it is still interesting to assess the initial spin since several mechanisms responsible for its growth after PBH formation have been explored~\cite{Jaraba:2021ces,Hofmann:2016yih,Calza:2021czr}.

The spin of the PBH can be estimated using the peak theory formalism~\cite{Yoo:2018kvb,DeLuca:2019buf} and is parameterized by the dimensionless Kerr parameter $a_*$. Its variance can be approximated by~\cite{Harada:2020pzb,Banerjee:2023qya}
\bea
\langle a_*^2 \rangle^{1/2} \simeq 4.01 \times 10^{-3} 
\frac{\sqrt{1-\gamma^2}}{1 + 0.036 \left[ 21 - 2 \log_{10} \left( \frac{f_{\rm PBH}}{10^{-7}} \right) - \log_{10} \left( \frac{M_{\rm PBH}}{10^{15} {\rm g}} \right) \right]} \, ,
\label{eq:spin}
\eea
and expressed in terms of the PBH mass and abundance. The parameter $\gamma$ is defined in terms of the first three spectral moments of the distribution of the curvature perturbations. Its deviation from unity describes the non-monochromaticity of such distribution. The parameter $\gamma$ typically ranges in $(0.85, 1)$~\cite{DeLuca:2019buf,Harada:2020pzb}. In the following we use a reference value of $\gamma \simeq 0.96$~\cite{Harada:2020pzb}.

\section{Numerical results}
\label{sec:results}
In this section, we present our numerical results for the GW signal and PBH fraction in the ALP model under consideration and point out the complementarity with existing ALP and PBH constraints.  
\subsection{GW-PBH complementarity}
\label{sec:5.1}
\begin{figure}[t!]
\centering
\includegraphics[width=\textwidth]{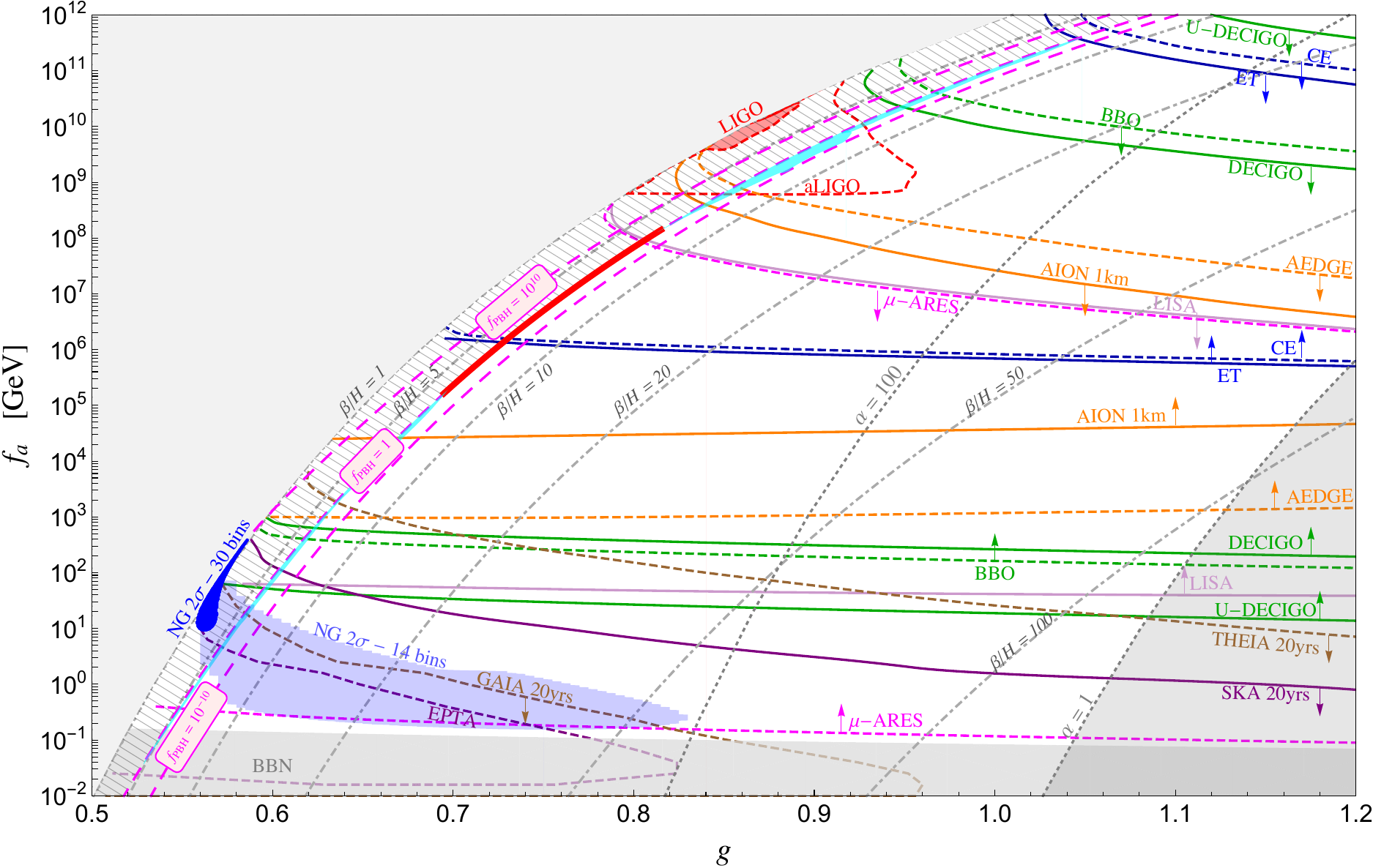}
\caption{PBH and GW predictions in our scale-invariant $SU(2)$ realization of the ALP model. Some contours of $f_{\rm PBH}$ (purple), $\alpha$ and $\beta/H$ are shown. The gray and red shaded regions are disfavored.  The thick red line shows the allowed region where PBHs can explain 100\% DM relic.  The blue shaded regions are the $2\sigma$ preferred regions to explain the NANOGrav signal. Other solid/dashed lines correspond to the sensitivities of the future GW detectors. See text for further details.} 
\label{fig:fa_g}
\end{figure}

In Fig.~\ref{fig:fa_g} we show the parameter space of the ALP model projected onto the plane defined by the gauge coupling constant $g$ and the ALP decay constant $f_a$. In this plane we show contour lines of several values of $\beta/H$. As already discussed above, in the gray region in the top left corner of the plot, the transition to global minimum never occurs and the system remains trapped in the inflating regime.

 We also show the contours of $\alpha=10^2$ and $\alpha = 1$, the latter approximately identifying the boundary of the region in which supercooling is not realized. In this portion of the parameter space (gray shaded region on the bottom right corner), totally irrelevant for the purpose of our work, our prediction for the GW spectra cannot be trusted as they all rely on the supercooling hypothesis.    

\begin{figure}[t!]
\centering
\includegraphics[width=\textwidth]{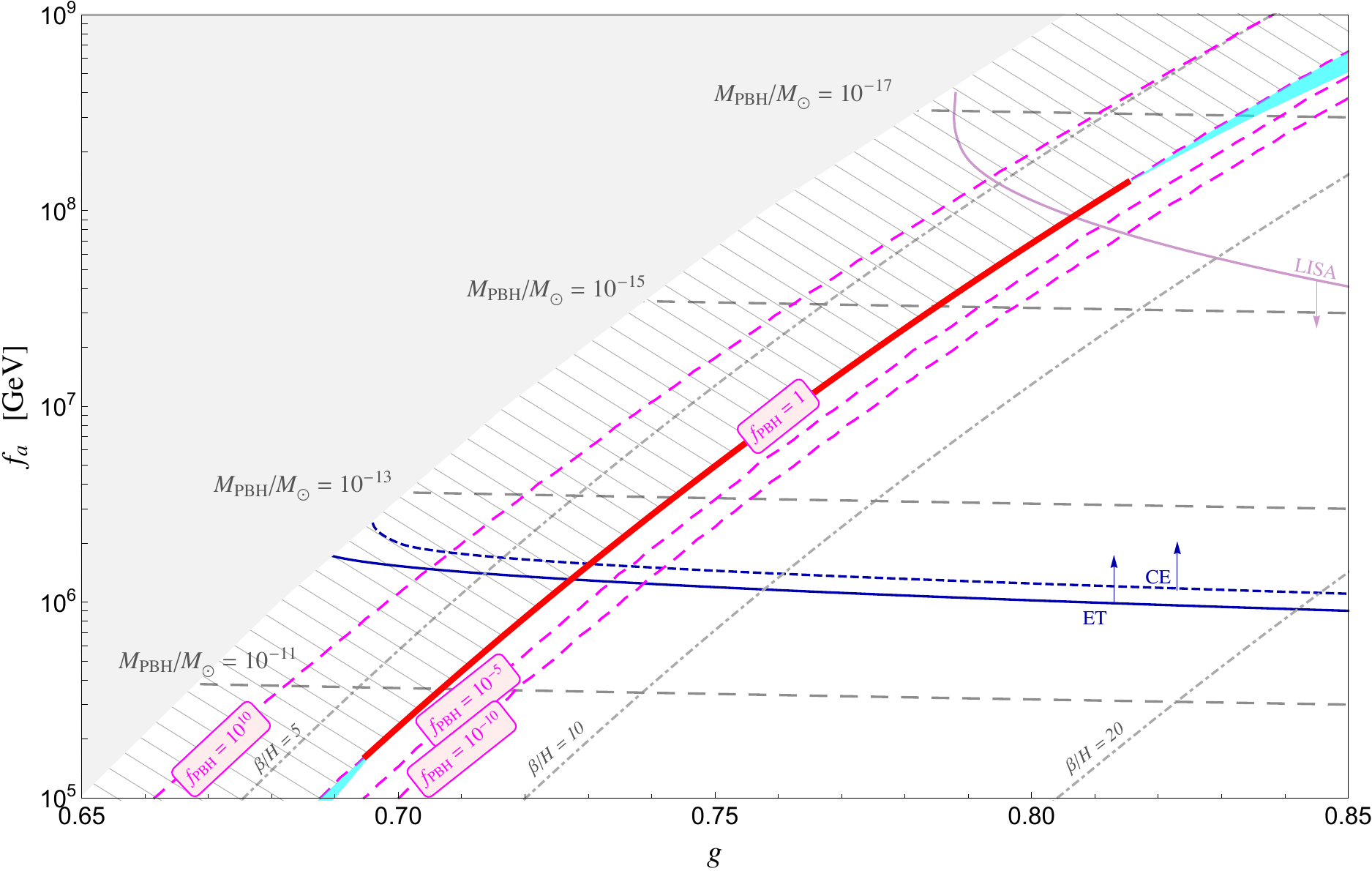}
\caption{Same as Fig.~\ref{fig:fa_g} but zoomed in around the $f_{\rm PBH}=1$ line. Also shown are some relevant $M_{\rm PBH}$ contours (in units of $M_\odot$).} 
\label{fig:fa_g_zoom}
\end{figure}

The three purple dashed lines delineate the values of $f_a$ and $g$ for which the fraction of DM in PBHs is equal to $f_\textrm{PBH} = 10^{10}, 1, 10^{-10}$, exemplifying, respectively, an overabundant PBH scenario, one in which the distribution of PBHs can exactly explains the DM relic, and a underabundant PBH case. The $f_{\rm PBH}=1$ case is characterized by values of $\beta/H \sim 5 - 7$, and the region above it is shown as hatched.  In the slice of the plot defined by $10^{-10} < f_\textrm{PBH} <1$, we report the constraints, given by shaded cyan regions, from the PBH searches that we will be detailed below. Concurrently, we highlight in red, on top of the $f_\textrm{PBH} = 1$ curve, the PBH mass window compatible with all the PBH bounds (see Section~\ref{sec:5.4}). This is realized by $\beta/H \sim 6.5 - 7$. See Fig.~\ref{fig:fa_g_zoom} for a zoomed-in view of this region. 

The horizontal gray band for $f_a \lesssim 100$ MeV is excluded by the requirement that, at the end of the PT, the plasma is reheated to a temperature well above the one at which BBN occurs. In particular, we enforce the conservative bound $T_{\rm RH} > 10$ MeV.

Finally, we consider the effective dark radiation bounds during BBN and CMB decoupling. In particular,
the energy density of the primordial GW has to be smaller than the limit on dark radiation encoded in $\Delta N_\text{eff.}$ from BBN and CMB observations since the gravitons behave as massless radiation degrees of freedom. Any change of the number of effective
relativistic degrees of freedom ($N_\text{eff}$) at the time of recombination is set by~\cite{Maggiore:1999vm}
\begin{align}
    \int_{f_\text{min}}^{\infty} \frac{\text{d}f}{f}   \Omega_\text{GW}(f) h^2 \leq 5.6\times10^{-6}\;\Delta N_\text{eff} \, .\label{eq:darkrad}
\end{align}
While the lower limit for the integration is $f_\text{min}\simeq 10^{-10}\text{Hz}$ for BBN and $f_\text{min}\simeq 10^{-18}\text{Hz}$ for the CMB bounds, in practice, when we plot \textit{e.g.} several GW spectra simultaneously, as a first-order estimate we may use the approximation to ignore the frequency dependence and to set bounds just on the energy density of the peak for a given GW spectrum; this is given by 
\begin{align}
    \Omega_\text{GW}^\text{Peak} h^2 \leq   5.6\times10^{-6}\;\Delta N_\text{eff} \, .\label{eq:darkrad2}
\end{align}
We consider the constraints on $\Delta N_{\rm eff}$ from BBN and the PLANCK 2018 limits~\cite{Planck:2018vyg}, as well as future reaches of CMB experiments such as CMB-S4~\cite{CMB-S4:2020lpa, CMB-S4:2022ght}, CMB-Bharat~\cite{CMBBharat:01} and CMB-HD~\cite{Sehgal:2019ewc,CMB-HD:2022bsz}. We find that the only relevant constraint can be enforced by CMB-HD around $\beta/H \lesssim 2$ and close to the separation boundary from the region in which the FOPT never completes.

In the plot, the expected sensitivity reaches of several current and future GW observatories are shown. These can be broadly classified as: 
\begin{itemize}
    \item \textbf{Ground-based  interferometer detectors:} a\textsc{LIGO}/a\textsc{VIRGO} (red dashed)~\cite{LIGOScientific:2014pky,VIRGO:2014yos,LIGOScientific:2019lzm}, \textsc{AION}~\cite{Badurina:2019hst} (orange solid), \textsc{Einstein Telescope (ET)}~\cite{Punturo:2010zz,Hild:2010id} (blue solid), \textsc{Cosmic Explorer (CE)} ~\cite{LIGOScientific:2016wof,Reitze:2019iox} (blue dashed). 
    \item   \textbf{Space-based interferometer detectors:}  \textsc{LISA}~\cite{LISA:2017pwj,Baker:2019nia} (pink solid), \textsc{BBO}~\cite{Crowder:2005nr,Corbin:2005ny} (green dashed), 
    \textsc{DECIGO}/\textsc{U-DECIGO}\cite{Yagi:2011wg,Kawamura:2020pcg} (green solid), \textsc{AEDGE}~\cite{AEDGE:2019nxb,Badurina:2021rgt} (orange dashed), \textsc{$\mu$-ARES}~\cite{Sesana:2019vho} (magenta dashed).
    \item \textbf{Recast from astrometry proposals (star surveys):} \textsc{GAIA}/\textsc{THEIA}~\cite{Garcia-Bellido:2021zgu} (brown dashed).  
     \item \textbf{Pulsar timing arrays:} \textsc{SKA}~\cite{Weltman:2018zrl} (purple), \textsc{EPTA}~\cite{Lentati:2015qwp,Babak:2015lua} (purple dashed),  \\ \textsc{NANOGrav}~\cite{NANOGrav:2023gor} (blue shaded region).
\end{itemize}
The only existing bound comes from the SGWB search by the LIGO-VIRGO collaboration~\cite{KAGRA:2021kbb}, as shown by the red shaded region. Also shown is the projected sensitivity at the end of the next phase, advanced LIGO-VIRGO~\cite{LIGOScientific:2014pky,VIRGO:2014yos,LIGOScientific:2019lzm}, which is already close to the region of the parameter space relevant for PBH production. The future Einstein Telescope and LISA interferometers will probe the space in which PBH from a supercooled FOPT can explain all the observed DM abundance.

In the bottom left corner of Fig.~\ref{fig:fa_g}, we show the 2$\sigma$ region (blue shaded) corresponding to the recent detection by the NANOGrav collaboration~\cite{NANOGrav:2023gor} under the hypothesis that the observed SGWB originates from a PT. The region is obtained by minimizing the follwoing figure of merit:
\bea
\chi^2 = \sum_i^N \frac{(\log_{10} \Omega_{\rm th} h^2 - \log_{10} \Omega_{\rm exp} h^2)^2}{\sigma_i^2} \, ,
\eea
where $\Omega_{\rm th} h^2$ and $\Omega_{\rm exp} h^2$ represent, respectively, the GW relic from
our theoretical prediction of the supercooled FOPT and the experimental value
from NANOGrav. For the estimate of the $\chi^2$ we ignore the uncertainty in the frequency width and just consider, for each bin, the uncertainty $\sigma_i$ in the $\log_{10}(\Omega_{\rm exp} h^2)$, which we approximate using the half-width.  The 2$\sigma$ region is determined in two ways: (i) using all 30 frequency bins as given by NANOGrav~\cite{NANOGrav:2023gor}, and (ii) using only the first 14 bins at lower frequencies.\footnote{Previous analyses~\cite{Ellis:2023dgf, Gouttenoire:2023bqy, Ellis:2023oxs} only considered the first 14 bins, presumably because of upward fluctuations in the  higher-frequency bins, which are also in tension with the BBN and Planck constraints (see Fig.~\ref{fig:NGbestfit}).}
In case (i), we see that the interpretation of the NANOGrav signal as a stochastic background of GW from a supercooled FOPT in our ALP model is disfavored since it falls within the PBH overclosure region. However for case (ii) considering only the first 14 bins, there is allowed parameter space with $f_{\rm PBH}\leq 1$. This is consistent with previous analyses using more sophisticated statistical analysis tools~\cite{Gouttenoire:2023bqy, Ellis:2023oxs}. 
%However, this is only true if we assume instantaneous reheating. If there is a preheating or prolonged reheating period, aided by PBH evaporation (see e.g. Ref.~\cite{RiajulHaque:2023cqe}), some (or most) of the PBH population can in principle be diluted away by the end of reheating to relax the overclosure bound and save the NANOGrav-preferred region. We postpone this discussion to a future work. 

In order to extract the sensitivity regions, we exploit the analysis developed in Ref.~\cite{Thrane:2013oya} in which, from the effective noise strain $S_\textrm{noise}$ provided by the experimental collaborations, we compute the power-law integrated (PLI) limit by maximizing the signal-to-noise (SNR) ratio over the spectral index. 
The SNR for a signal $\Omega(f_{\rm gw})$ is defined as
\bea
\mathrm{SNR}=\sqrt{t_{\rm obs} \int_{f_{\rm min}}^{f_{\rm max}} df \bigg[\frac{\Omega(f)}{\Omega_{\rm noise}(f)}\bigg]^2}\,,   \qquad   \Omega_{\rm noise}(f_{\rm gw})=\frac{2\pi^2}{3H_0^2} f^3 S_{\rm noise}(f_{\rm gw})\,,
\eea
where the prefactor $t_{\rm obs}$ represents the integrated observational time, multiplied by the number of interferometers involved in the experiment. To determine a conservative bound, we assume a power-law family of signals $\Omega_b(f_{\rm gw})=A_b f_{\rm gw}^b$ and we extract, at each frequency $f_{\rm gw}$, the largest value of $\Omega_b(f_{\rm gw})$ compatible with a given reference value of $\mathrm{SNR}_{\rm ref}$, here taken as 10. This gives the (PLI) limit
\bea
\Omega_{\rm PLI}(f_{\rm gw}) &=& \max_b\,\, \Omega_b(f_{\rm gw})\big|_{\rm SNR_{ref}}= \max_b\,\, A_b\big|_{\rm SNR_{ref}} f_{\rm gw}^b \nonumber \\
&=& \frac{\mathrm{SNR}_{\rm ref}}{\sqrt{ t_{\rm obs}}} \max_b  \bigg[\bigg(\int_{f_{\rm min}}^{f_{\rm max}} d\bar f \frac{\bar f^{2b}}{\Omega_{\rm noise}^2(\bar f)}\bigg)^{-\frac{1}{2}}\, f_{\rm gw}^b\bigg] \,.
\eea

In Fig.~\ref{fig:fa_g_zoom} we show a zoomed-in version of Fig.~\ref{fig:fa_g} around the region in which the distribution of PBH can explain the total DM abundance ($f_{\rm PBH}=1$), thus emphasizing the corresponding variability range of the two microsphysical parameters, $g$ and $f_a$. Dashed gray lines show some representative values of the PBH mass which give the correct DM relic density. As mentioned above, future GW detectors like LISA, AION, ET and CE will be able to probe this entire allowed parameter space in our ALP model.

\subsection{The astrophysical foreground}
\begin{figure}[t!]
\centering
\includegraphics[width=\textwidth]{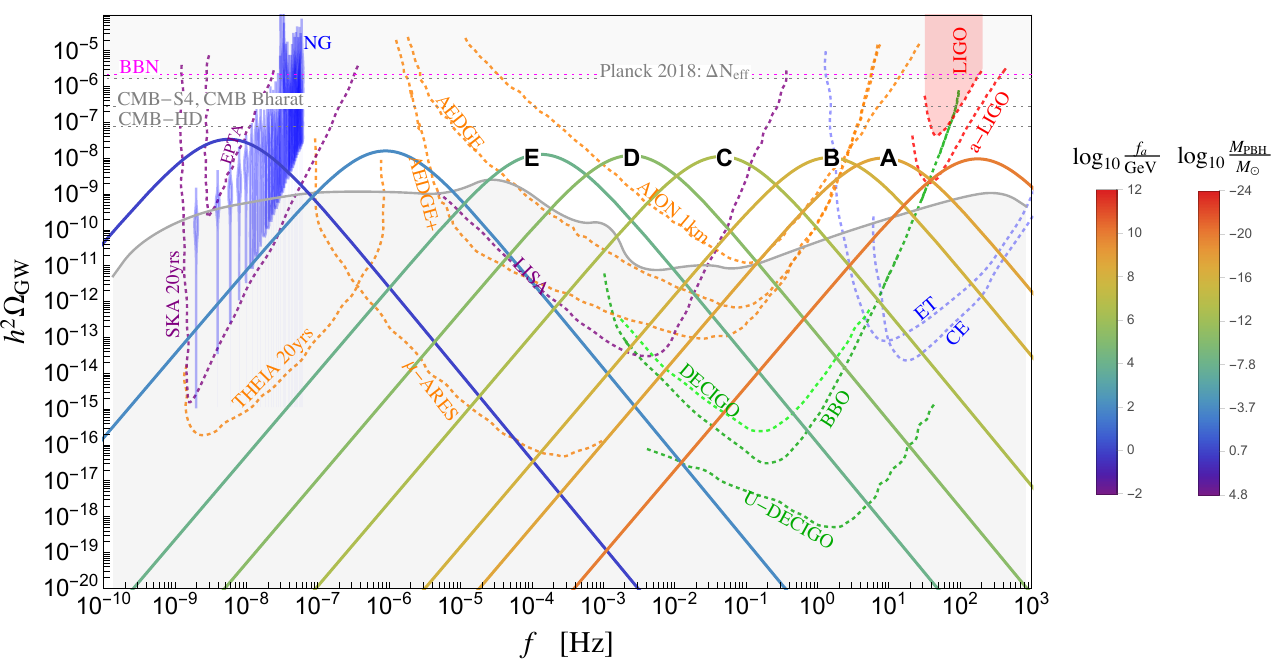}
\caption{The GW spectra predicted in our ALP model are shown by the colored solid lines corresponding to different values of $f_a$, which have a one-to-one correspondence with the PBH mass (for a fixed $f_{\rm PBH}=1$). Five benchmark points (A to E) are marked here, which will be used later. The areas above the colored dotted lines correspond to the projected sensitivities for several current and future GW observatories. The horizontal lines show current $\Delta N_{\rm eff}$ constraints from BBN and Planck data (shaded regions) and future reaches by CMB-S4, CMB-Bharat and CMB-HD. The red shaded region on top right is the current exclusion from LIGO-VIRGO data. The blue violins on the top left are the recent NANOGrav data points for the SGWB detection. The shaded region on the bottom half of the plane is the expected astrophysical foreground. See text for details.} 
\label{fig:GW}
\end{figure}

As mentioned in the Introduction, the cosmological stochastic GW signal discussed here is subject to an astrophysical foreground from unresolved compact binary coalescences~\cite{LIGOScientific:2016fpe}. LIGO/VIRGO has already observed GW events involving binary black hole (BH-BH) and binary neutron stars (NS-NS) mergers~\cite{LIGOScientific:2018mvr, Nitz:2021zwj}; hence, an astrophysical foreground is guaranteed to exist. The sum of the diffuse astrophysical foreground is shown in Fig.~\ref{fig:GW} as the gray shaded region in the bottom half plane (see also Refs.~\cite{Dev:2016feu, Ghoshal:2023pcx}). 
Different subtraction methods have been proposed to deal with this foreground at future GW detectors and to extract any potential signal of primordial origin~\cite{Cutler:2005qq, Zhu:2012xw, Regimbau:2016ike, Sachdev:2020bkk, Zhou:2022nmt,Zhong:2022ylh,Pan:2023naq,Bellie:2023jlq, Song:2024pnk}. We expect that the NS and BH foreground can be subtracted from the sensitivities of BBO and ET or CE in the ranges $\Omega_{\rm GW} \simeq 10^{-15}$~\cite{Cutler:2005qq} and $\Omega_{\rm GW} \simeq 10^{-13}$~\cite{Regimbau:2016ike}.
However in the LISA window, the galactic and extra-galactic binary white dwarf  (WD-WD) foregrounds may dominate over the NS-NS and BH-BH foregrounds~\cite{Farmer:2003pa, Rosado:2011kv, Moore:2014lga} and should be subtracted~\cite{Kosenko:1998mv} from the expected sensitivity $\Omega_{\rm GW} \simeq 10^{-13}$ to be reached at LISA~\cite{Adams:2010vc, Adams:2013qma}. Given that such subtractions can be made possible with a network of GW detectors due to the fact that the GW spectrum generated by the astrophysical foreground increases with frequency as $f^{2/3}$~\cite{Wu:2011ac,Zhu:2012xw}, which is different from the GW spectrum generated by nucleating bubbles during strong FOPT ($\Omega_{\rm GW} \sim f^{3}$ and $\Omega_{\rm GW} \sim f^{-2.4}$ for frequencies below and above the peak, respectively; see Fig.~\ref{fig:GW}), we hope to be able to eventually pin down the GW signals from supercooled PT.

\subsection{Fitting the NANOGrav signal} \label{sec:5.3}
\begin{figure}[t!]
\centering
\includegraphics[width=\textwidth]{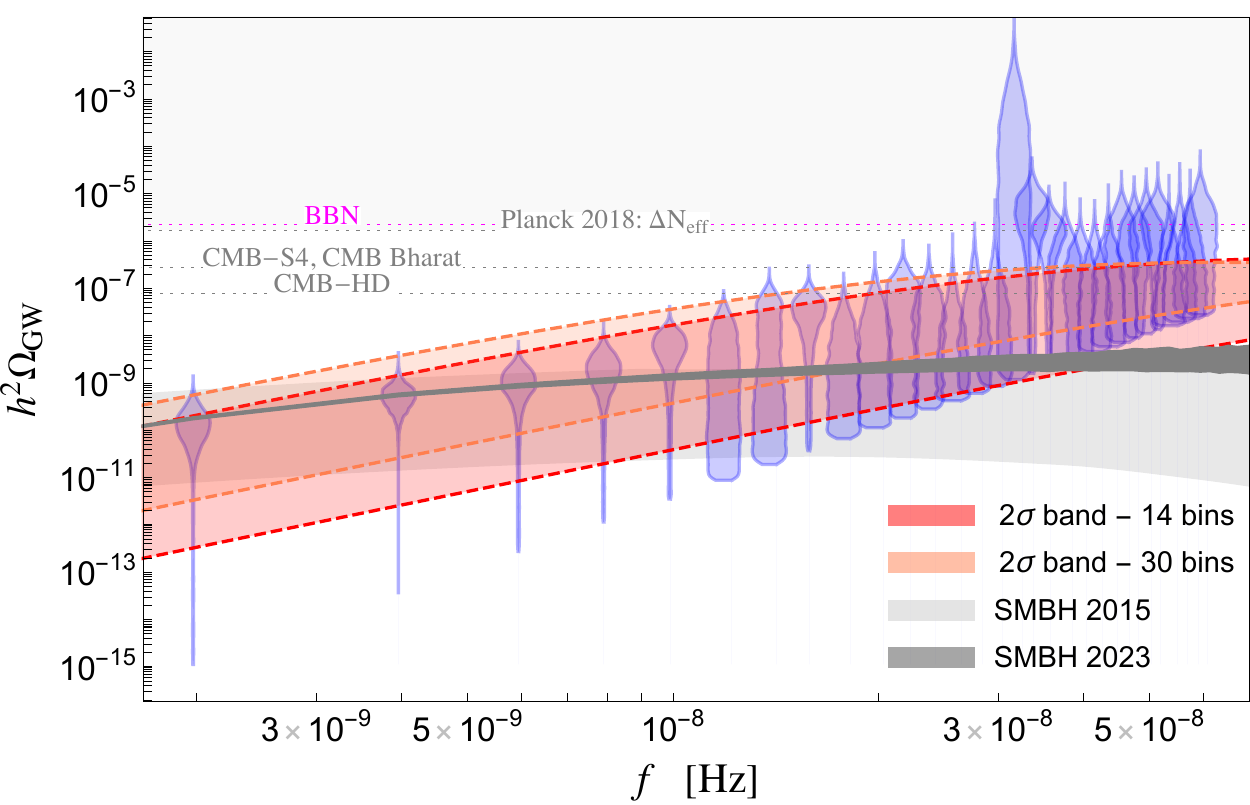}
\caption{The 15-year data from the NANOGrav collaboration (blue violins) and the $2\sigma$ bands (shown by the orange and red shaded regions) corresponding to the best-fit GW spectrum from the supercooled FOPT in our ALP model. For comparison, the expected astrophysical contribution from SMBH is shown by the two gray bands. Other labels are the same as in Fig.~\ref{fig:GW}. 
}
\label{fig:NGbestfit}
\end{figure}

In Fig.~\ref{fig:NGbestfit}, we focus on the nHz frequency regime and the 15-year NANOGrav data on the SGWB~\cite{NANOGrav:2023gor}. Our ALP model can explain this signal for $f_a\simeq (10~{\rm GeV}-1~{\rm TeV})$ and $g\simeq 0.56-0.60$. As explained in the context of Fig.~\ref{fig:fa_g}, we calculate the $2\sigma$ band that fits the NANOGrav signal in two ways, i.e. with only the first 14 bins (red shaded) and including all 30 bins (orange shaded). For comparison, we also show the expected band of spectra from astrophysical sources, namely, supermassive black hole (SMBH) binaries as evaluated in~\cite{Rosado:2015epa} and, more recently, in~\cite{NANOGrav:2023hfp}. Our cosmological signal tends to fit the NANOGrav data slightly better, especially if we take all 30 bins. Our $2\sigma$ best-fit spectra are also consistent with the $\Delta N_{\rm eff}$ bound from BBN and Planck data (horizontal gray shaded region), although the NANOGrav data in the high-frequency bins are in some tension with the BBN and Planck constraints. 

\subsection{PBH constraints} \label{sec:5.4}
%%%%
\begin{figure}[t!]
\centering
\includegraphics[width=\textwidth]{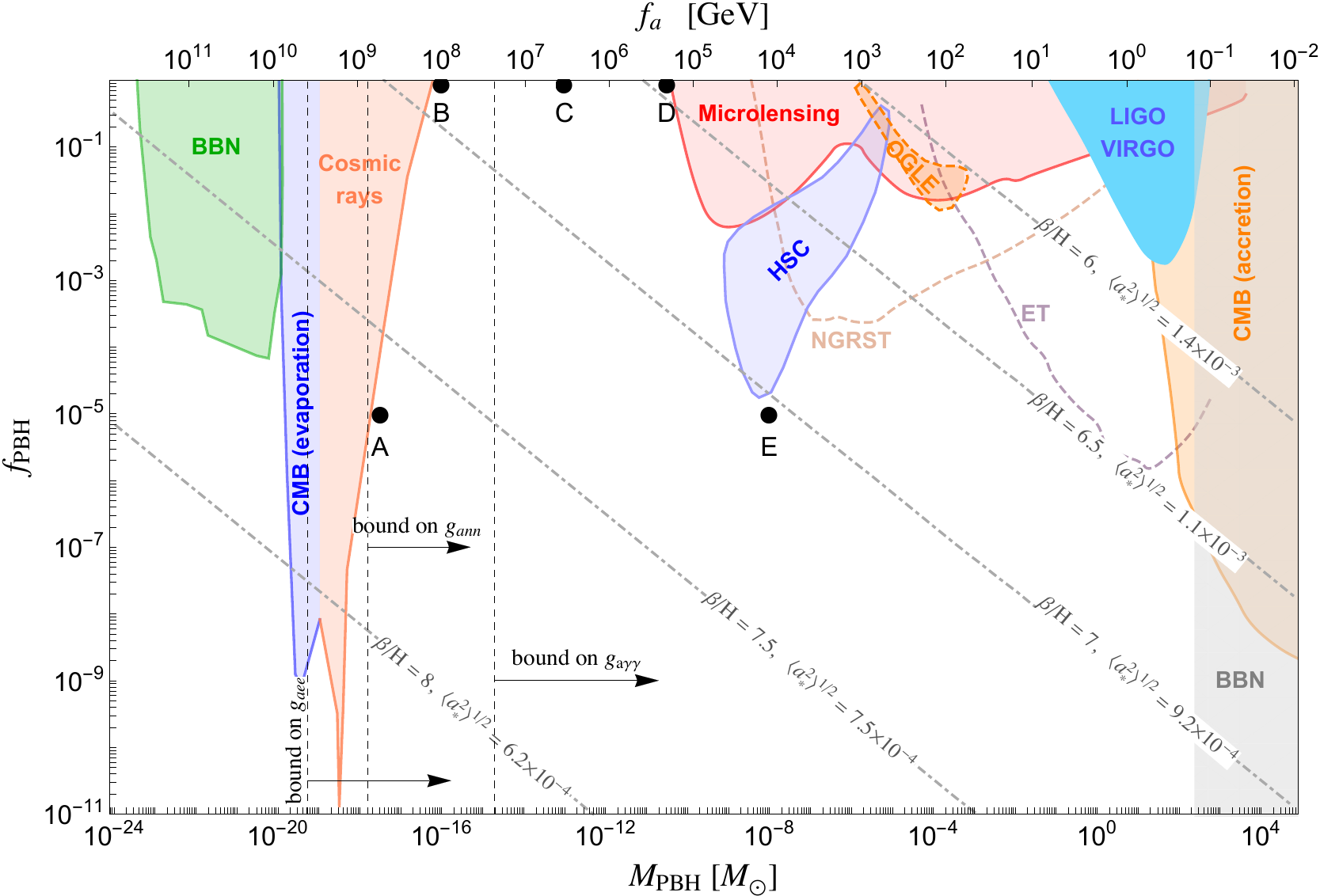}
\caption{Summary of the PBH constraints. The colored shaded areas are excluded by BBN, CMB, cosmic rays, microlensing and GW observations, as discussed in the text. The future sensitivities of NGRST and ET are shown by the dashed curves. We also depict the initial PBH spin $a_{*}$ that is formed. The ALP decay constant value corresponding to each PBH mass is shown on the upper $x$-axis. We also show the stringent experimental constraints from complementary ALP searches (vertical lines) if ALP couples to photons, electrons or nucleons (see Section~\ref{sec:complementarity}). The gray shaded region on the right edge with $f_a\lesssim 100$ MeV is also excluded by BBN. 
}
\label{fig:pbh}
\end{figure}

In Fig.~\ref{fig:pbh} we depict the observational constraints on $f_{\rm PBH}$ (see Ref.~\cite{Green:2020jor} for a review). PBH evaporation via Hawking radiation give constraints in (see also Refs.~\cite{Saha:2021pqf,Laha:2019ssq,Ray:2021mxu}): CMB~\cite{Clark:2016nst}, EDGES~\cite{Mittal:2021egv},  INTEGRAL~\cite{Laha:2020ivk,Berteaud:2022tws}, Voyager~\cite{Boudaud:2018hqb}, 511\;keV~\cite{DeRocco:2019fjq,Dasgupta:2019cae}, and 
EGRB~\cite{Carr:2009jm}. The microlensing-related constraints from HSC (Hyper-Supreme Camera)~\cite{Niikura:2017zjd}, EROS~\cite{EROS-2:2006ryy} and Icarus~\cite{Oguri:2017ock}, as well as the hint of PBH from OGLE~\cite{Niikura:2019kqi} are also shown.  We also show the future micro-lensing sensitivity reach of Nancy Grace Roman Space Telescope (NGRST)~\cite{DeRocco:2023gde}. Constraints coming from modification of the CMB spectrum due to PBH accretion are shown on the right~\cite{Serpico:2020ehh} (see also Ref.~\cite{Piga:2022ysp}). Finally the mass range around $M_{\odot}$ is constrained by LIGO-VIRGO observations on PBH-PBH merger ~\cite{Franciolini:2022tfm,Kavanagh:2018ggo,Hall:2020daa,Wong:2020yig,Hutsi:2020sol,DeLuca:2021wjr,Franciolini:2021tla}, while future GW interferometers like ET will be able to set better limits on the PBH abundance~\cite{DeLuca:2021hde,Pujolas:2021yaw,Franciolini:2022htd,Martinelli:2022elq,Franciolini:2023opt,Branchesi:2023mws} as depicted by the dashed gray curve in the plot. 

Due to the one-to-one correspondence between the PBH mass and the ALP decay constant, we also show in Fig.~\ref{fig:pbh} the $f_a$ values on the upper $x$-axis. This brings in additional constraints. For instance, the region on the right with $f_a\lesssim 100$ MeV, which roughly corresponds to $T_{\rm RH}\lesssim 10$ MeV, is excluded by general BBN considerations. Similarly, if the ALP couples to SM photons, electrons or nucleons, stringent laboratory constraints apply on the corresponding couplings $g_{a\gamma\gamma}, g_{aee}$ and $g_{ann}$, respectively (see Section~\ref{sec:complementarity}). Since these couplings scale as $f_a^{-1}$ (with ${\cal O}(1)$ coefficients), upper limits on these couplings can be translated into lower limits on the $f_a$ scale, or upper limits on the PBH mass, as shown by the vertical lines.     

Given the PBH constraints in Fig.~\ref{fig:pbh}, we take five benchmark points A, B, C, D, E in the allowed region, with the masses of the PBHs formed during the supercooling as  $10^{-17.5} M_{\odot}$, $10^{-16} M_{\odot}$, $10^{-13}M_{\odot}$, $10^{-10.5}M_{\odot}$ and $10^{-8}M_{\odot}$ respectively. While B, C and D correspond to the PBH accounting for the entire DM content of the Universe, for benchmark points A and E, PBHs only comprise  
$10^{-5}$ of the DM relic density of the Universe. The GW spectra for these benchmark points are shown in Fig.~\ref{fig:GW}. 

The contours of the initial PBH spin $a_*$ that is formed during supercooling [cf.~Eq.~\eqref{eq:spin}] is also shown in Fig.~\ref{fig:pbh}, along with the corresponding $\beta/H$ values. It is clear that the spin is very small, although nonzero. It is unlikely that such a tiny spin can be observed (e.g. via superradiance). 

The five benchmark points chosen in Fig.~\ref{fig:pbh} correspond to PBHs of different masses and abundances. Points B, C, D are with $f_{\rm PBH}=1$, whereas A and E are with $f_{\rm PBH}=10^{-5}$. Moreover, they all correspond to different initial spin. Also, as shown in Fig.~\ref{fig:GW}, different GW experiments at different frequencies are sensitive to these benchmarks.

For axion decay constant values $f_a = 10^5 - 10^8$ GeV, PBH can be the entire DM candidate of the universe (see Fig. 7) while for other values, PBH can still be the fractional DM candidate while the dominant component of the DM may come from ALP relic produced due to vacuum misalignment mechanism~\cite{DiLuzio:2020wdo}, see Fig. 8.

\medskip

\section{Complementarity with laboratory and astrophysical ALP searches}
\label{sec:complementarity}

\begin{figure}[t!]
\centering
\includegraphics[width=\textwidth]{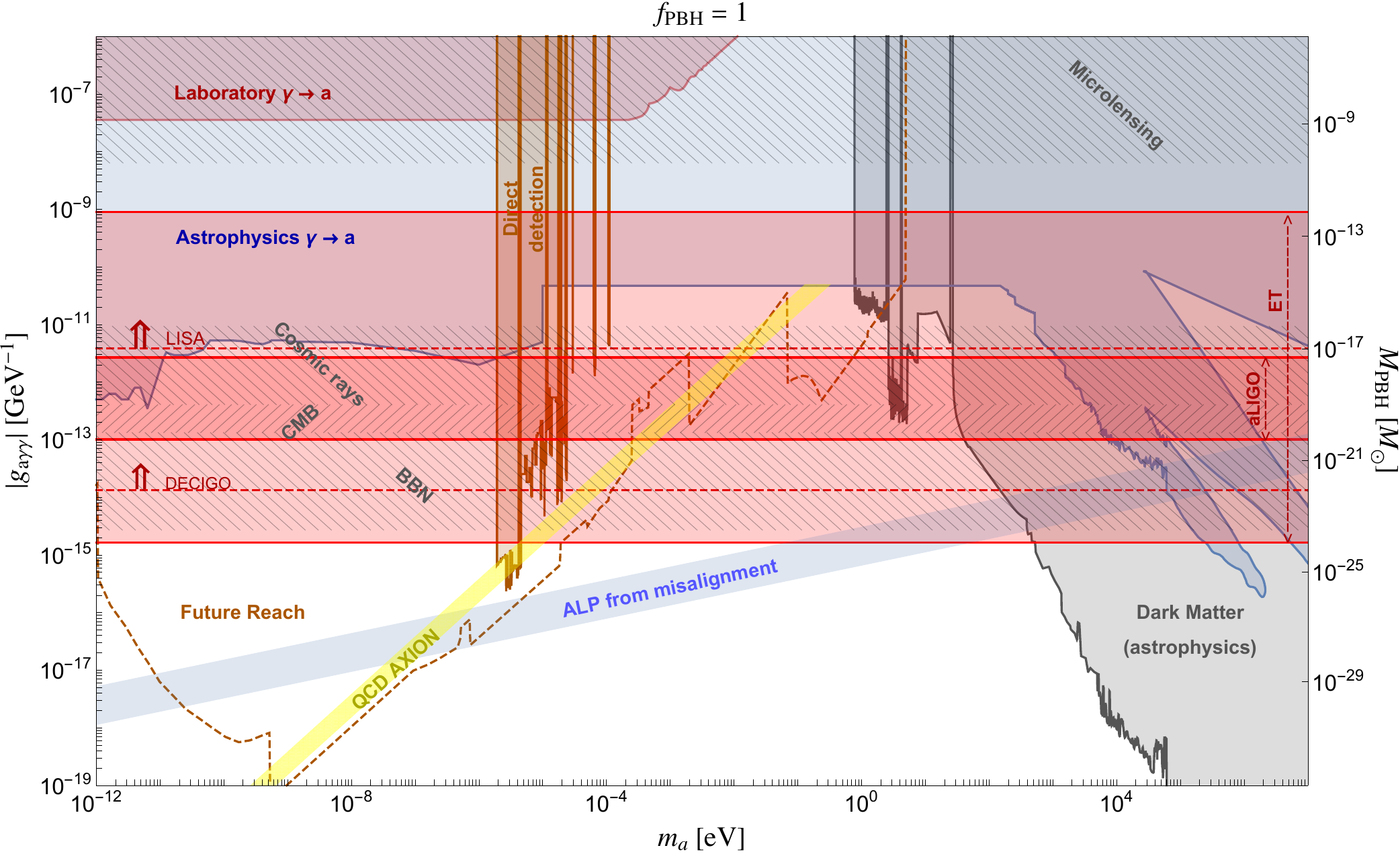}
\caption{Complementarity between PBH observations, the future GW sensitivities, and constraints on ALP-photon coupling $g_{a\gamma\gamma}$ from laboratory, astrophysical and
cosmological constraints. The GW prospects are shown by the 
red dotted lines. The PBH constraints are shown by the hatched regions. Other available ALP constraints are shown by the shaded regions. 
See text for more details. 
} 
\label{fig:axionphoton_fPBH1}
\end{figure}

\begin{figure}[t!]
\centering
\includegraphics[width=\textwidth]{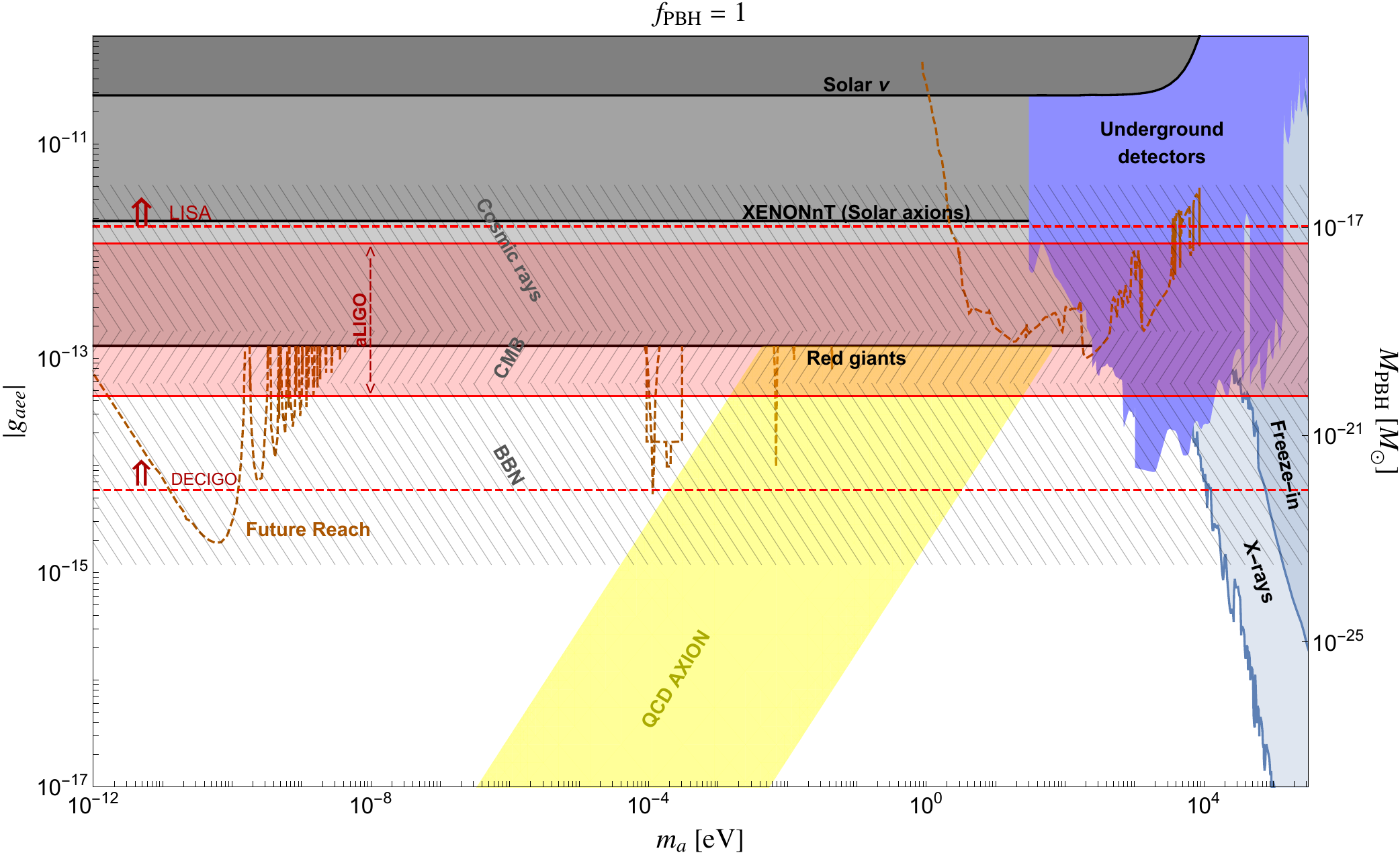}
\caption{Similar to Fig.~\ref{fig:axionphoton_fPBH1} but for the ALP coupling to electrons ($g_{aee}$). } 
\label{fig:axionelectron_fPBH1}
\end{figure}

\begin{figure}[t!]
\centering
\includegraphics[width=\textwidth]{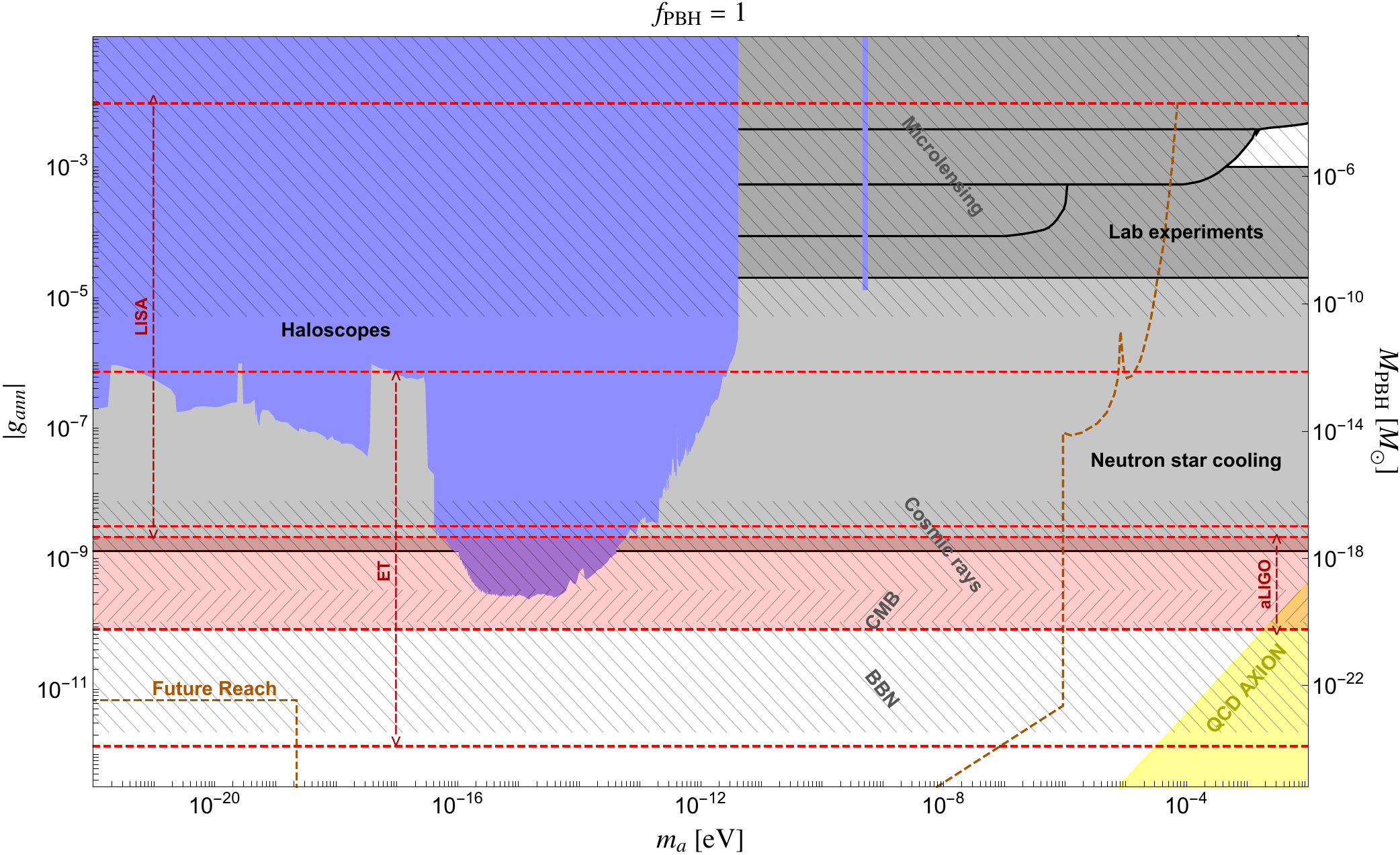}
\caption{Similar to Fig.~\ref{fig:axionphoton_fPBH1} but for thr ALP coupling to nucleons ($g_{ann}$). } 
\label{fig:axionneutron_fPBH1}
\end{figure}

At low energies, the ALP couplings to the SM particles can be induced by higher-dimensional operators. The lowest-order dimension-5 couplings are proportional to the inverse power of $f_a$. At this order, the effective couplings of the ALP $a$ to the SM photon and fermions can be written as 
\begin{align}
    {\cal L}_a \supset -\frac{g_{a\gamma\gamma}}{4}a F_{\mu\nu}\tilde{F}^{\mu\nu}-a\sum_f g_{aff}(i\bar f \gamma^5 f) \, ,
\end{align}
where $g_{a\gamma\gamma}\simeq \alpha_{\rm em}/(2\pi f_a)$ and $g_{aff}\simeq m_f/f_a$, neglecting ${\cal O}(1)$ coefficients. From Figs.~\ref{fig:fa_g} and \ref{fig:GW}, we see that the GW experiments constrain the scale $f_a$ independently of the ALP mass. Therefore, the GW limits can be translated to limits on the couplings $g_{a\gamma\gamma}$, $g_{aee}$ and $g_{ann}$ for photons, electrons and nucleons respectively~\cite{Dev:2019njv}, which are typically used for laboratory and astrophysical searches of ALPs~\cite{AxionLimits}. Moreover, since $f_a$ is directly correlated with the PBH mass  in our ALP model, the constraints on $M_{\rm PBH}$ from Fig.~\ref{fig:pbh} can be translated into constraints on $f_a$, which in turn can be translated into constraints on the couplings $g_{a\gamma\gamma}$, $g_{aee}$ and $g_{ann}$. Thus, we get the three-prong complementarity between GW experiments, PBH fraction and ALP searches. This is depicted in Figs.~\ref{fig:axionphoton_fPBH1}, \ref{fig:axionelectron_fPBH1} and \ref{fig:axionneutron_fPBH1} for  $g_{a\gamma\gamma}$, $g_{aee}$ and $g_{ann}$, respectively with the choice $f_{\rm PBH}=1$. Analogous plots with $f_{\rm PBH}=10^{-5}$ are given in Appendix~\ref{ref:supp}.  

In Fig.~\ref{fig:axionphoton_fPBH1}, different shaded regions correspond to the laboratory constraints (dark red) from light-shining-through-wall (e.g.,~ALPS~\cite{Ehret:2010mh}, CROWS~\cite{Betz:2013dza}, OSQAR~\cite{OSQAR:2015qdv}), helioscope (e.g.,~CAST~\cite{CAST:2017uph}), and haloscope (e.g.,~ADMX~\cite{ADMX:2021nhd}) experiments,  as well as astrophysical (blue) constraints from SN1987A~\cite{Caputo:2022mah,Diamond:2023scc,Muller:2023vjm}, globular clusters~\cite{Dolan:2022kul}, and pulsar polar cap~\cite{Noordhuis:2022ljw} on the ALP-photon coupling. Assuming ALP to be the DM, further astrophysical/cosmological constraints are obtained from late-time ALP decays to photons (from EBL, $X$-rays, $\gamma$-rays, etc)~\cite{Cadamuro:2011fd, Foster:2021ngm} as shown by the bottom-right gray-shaded region. For details, see Refs.~\cite{AxionLimits, Choi:2020rgn}. The blue band at the bottom corresponds to the region where ALPs can reproduce the correct DM relic density from the misalignment mechanism~\cite{Arias:2012az, Adams:2022pbo}. This is especially relevant when the PBH by itself cannot explain all the DM fraction, and the ALP can serve as the remaining DM component. The yellow band is where the QCD axion lives (since $m_af_a\approx m_\pi f_\pi$)~\cite{Gorghetto:2018ocs}.  The dashed curve extending to the bottom shows the projected experimental reach from a combination of future helioscopes, haloscopes and other laboratory experiments~\cite{AxionLimits}. 

The GW prospects for $g_{a\gamma\gamma}$ are shown by the red dotted lines for future detectors like advanced-LIGO, ET, LISA and DECIGO. We see that the absence of a SGWB at these experiments can constrain additional ALP parameter space beyond the existing limits. More importantly, the {\em current} constraints on $M_{\rm PBH}$ from BBN, CMB, cosmic rays and microlensing [cf.~Fig.~\ref{fig:pbh}] are shown here by the hatched regions, some of which turn out to be the most stringent constraints in this parameter space. This is what we call `slaying ALPs with PBHs'. Of course, the PBH constraints become weaker if we take $f_{\rm PBH}<1$ (see Appendix~\ref{ref:supp}).  

Similarly, Fig.~\ref{fig:axionelectron_fPBH1} shows the constraints for the ALP-electron coupling. Here the laboratory constraints include those from underground detectors (blue shaded) such as XENON1T~\cite{XENON:2020rca,XENON:2021qze}, XENONnT~\cite{XENON:2022ltv}, PandaX~\cite{PandaX:2017ock}, DarkSide~\cite{DarkSide:2022knj}, EDELWEISS~\cite{EDELWEISS:2018tde}, SuperCDMS~\cite{SuperCDMS:2019jxx}, and GERDA~\cite{GERDA:2020emj}. The astrophysical constraints (gray/light-blue shaded) include those from solar neutrinos~\cite{Gondolo:2008dd}, red giant branch~\cite{Capozzi:2020cbu}, $X$-rays from ALP DM decays~\cite{Ferreira:2022egk} and freeze-in~\cite{Langhoff:2022bij}. The future projections include novel experiments with NV centers~\cite{Chigusa:2023hms} and magnons~\cite{Chigusa:2020gfs, Mitridate:2020kly, Ikeda:2021mlv}. We again find that the current PBH constraints are actually already ruling out most of the parameter space that could be probed by future laboratory searches, highlighting the usefulness of the complementarity approach discussed here. 

Finally, in Fig.~\ref{fig:axionneutron_fPBH1} we show the constraints on the ALP-nucleon coupling. In this case, the existing haloscope constraints (blue shaded), coming from experiments like CASPEr~\cite{JacksonKimball:2017elr}, K-$^3$He comagnetometer~\cite{Lee:2022vvb}, nEDM~\cite{Abel:2017rtm}, NASDUCK~\cite{Bloch:2021vnn, Bloch:2022kjm}, JEDI~\cite{JEDI:2022hxa} and ChangE~\cite{Wei:2023rzs, Xu:2023vfn}, and astrophysical constraints from isolated neutron star cooling~\cite{Buschmann:2021juv} are relevant. But the PBH constraints from BBN and CMB turn out to be the strongest, even cutting into the future laboratory sensitivities.

\section{Discussion and Conclusion}
\label{sec:conclusions}
%%%%%%%%%
In summary, we investigated the GW signal and the formation of PBHs during strong FOPT in an approximately-conformal ALP model. 
This model already contains a cosmologically stable DM candidate in the form of the ALP, which is basically the phase of the scalar which breaks the $U(1)$ global symmetry via the misalignment mechanism. However, the main point of this work is that a strong FOPT is automatically realized in this scenario due to supercooling from radiative symmetry breaking of the close-to-conformal scalar potential, which not only gives potentially observable stochastic GW signals in future GW experiments, but also leads to the formation of PBHs that could account for a fraction (or whole) of the DM relic density. 

Furthermore, in our model there is a one-to-one correspondence between the ALP decay constant $f_a$ and the PBH mass $M_{\rm PBH}$. This leads to a three-prong complementarity between the GW signal, PBH signals and other laboratory/astrophysical ALP probes. In particular, depending on the ALP couplings to the SM particles, the ALP constraints can be translated into a lower bound on the PBH mass, independent of $f_{\rm PBH}$ (see Fig.~\ref{fig:pbh}). Analogously, we can use the existing PBH constraints to `slay' additional ALP parameter space, which for $f_{\rm PBH}=1$ turn out to be even more stringent than the future laboratory searches (see Figs.~\ref{fig:axionphoton_fPBH1}, \ref{fig:axionelectron_fPBH1} and \ref{fig:axionneutron_fPBH1}). This $f_{\rm PBH}=1$ region can be tested via the GW signal in future GW detectors like LISA, ET and CE (see Fig.~\ref{fig:fa_g_zoom}). We also discussed the recent NANOGrav detection of a SGWB and its possible interpretation in terms of the FOPT in our ALP scenario (see Fig.~\ref{fig:NGbestfit}).  

Note that models of slow roll inflation producing PBHs also result in detectable GWs, in the same frequency range as from the FOPT, due to the enhanced power spectrum on small scales leading to significant anisotropic stress~\cite{Nakama:2016gzw,Garcia-Bellido:2017aan,Cai:2018dig,Bartolo:2018evs,Bartolo:2018rku,Qin:2023lgo} or with multiple fields beyond slow roll~\cite{Chen:2023lou}. However, the GW spectral shapes arising in those scenarios involving second-order tensor perturbations (also known as scalar-induced GW) are quite different from those of PTs, and this could be easily utilized to distinguish between different PBH formation mechanisms.

In future it would be interesting to consider a detailed inflationary scenario which involves modification to the  power spectrum on small scales due to the PT, especially in the critical Hubble patches, that would also induce anisotropic stress and a further GW source at second-order, and to study the GW spectral shape features in totality. Due to presence of spectator fields during inflation which undergo PT it may be natural to expect the GW signals observed to be non-Gaussian and to carry features of the primordial spectrum \cite{Kumar:2021ffi}. 
Also, it would be interesting to go beyond the expected PBH spectrum (beyond the monochromatic approximation used here) from the bubble collision, together with any additional GWs at second order in perturbation theory, in light of testing the $f_{\rm PBH}=1$ region at upcoming GW interferometer-based experiments.

\section*{Acknowledgments}
We thank Yann Gouttenoire, Kayhan Gultekin, Marek Lewicki, Priyamvada Natarajan and Kai Schmitz for comments on the draft. 
A.C. thanks the Galileo Galilei Institute for Theoretical Physics for the hospitality and the INFN for partial support during the completion of this work.
This work has been partly funded by the European Union – Next Generation EU through the research grant number P2022Z4P4B “SOPHYA - Sustainable Optimised PHYsics Algorithms: fundamental physics to build an advanced society" under the program PRIN 2022 PNRR of the Italian Ministero dell’Università e Ricerca (MUR) and partially supported by ICSC – Centro Nazionale di Ricerca in High Performance Computing, Big Data and Quantum Computing. The work of B.D. was partly supported by the U.S. Department of Energy under grant No. DE-SC 0017987.

\appendix
\section{Supplementary plots}
\label{ref:supp}
\begin{figure}[t!]
\centering
\includegraphics[width=\textwidth]{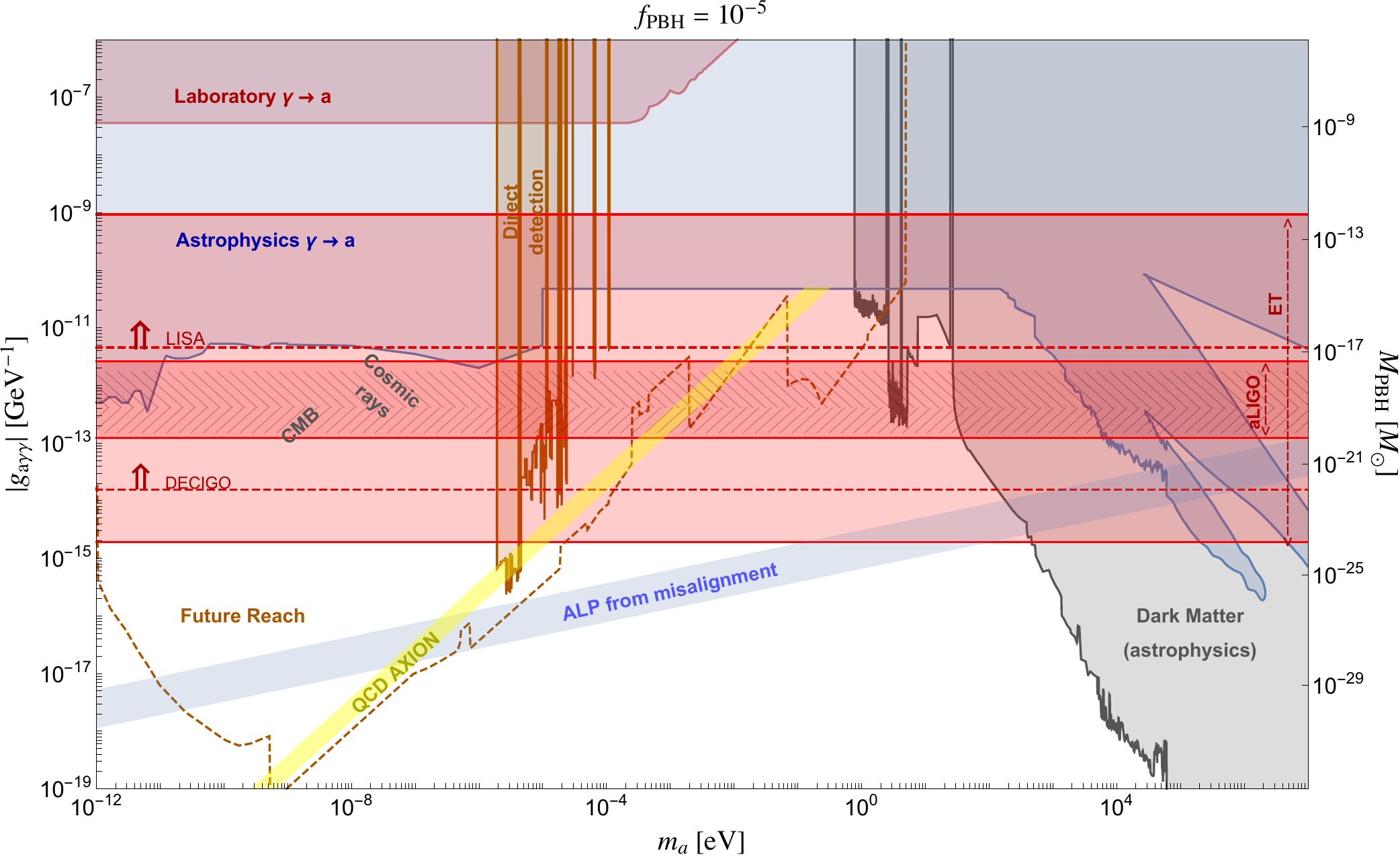}
%\vspace*{-0.5cm}
\caption{Same as Fig.~\ref{fig:axionphoton_fPBH1} but for $f_{\rm PBH} = 10^{-5}$. Only the PBH bounds become weaker, while all other bounds remain the same. } 
\label{fig:axionphoton_fPBHm5}
\end{figure}

\begin{figure}[h!]
\centering
\includegraphics[width=\textwidth]{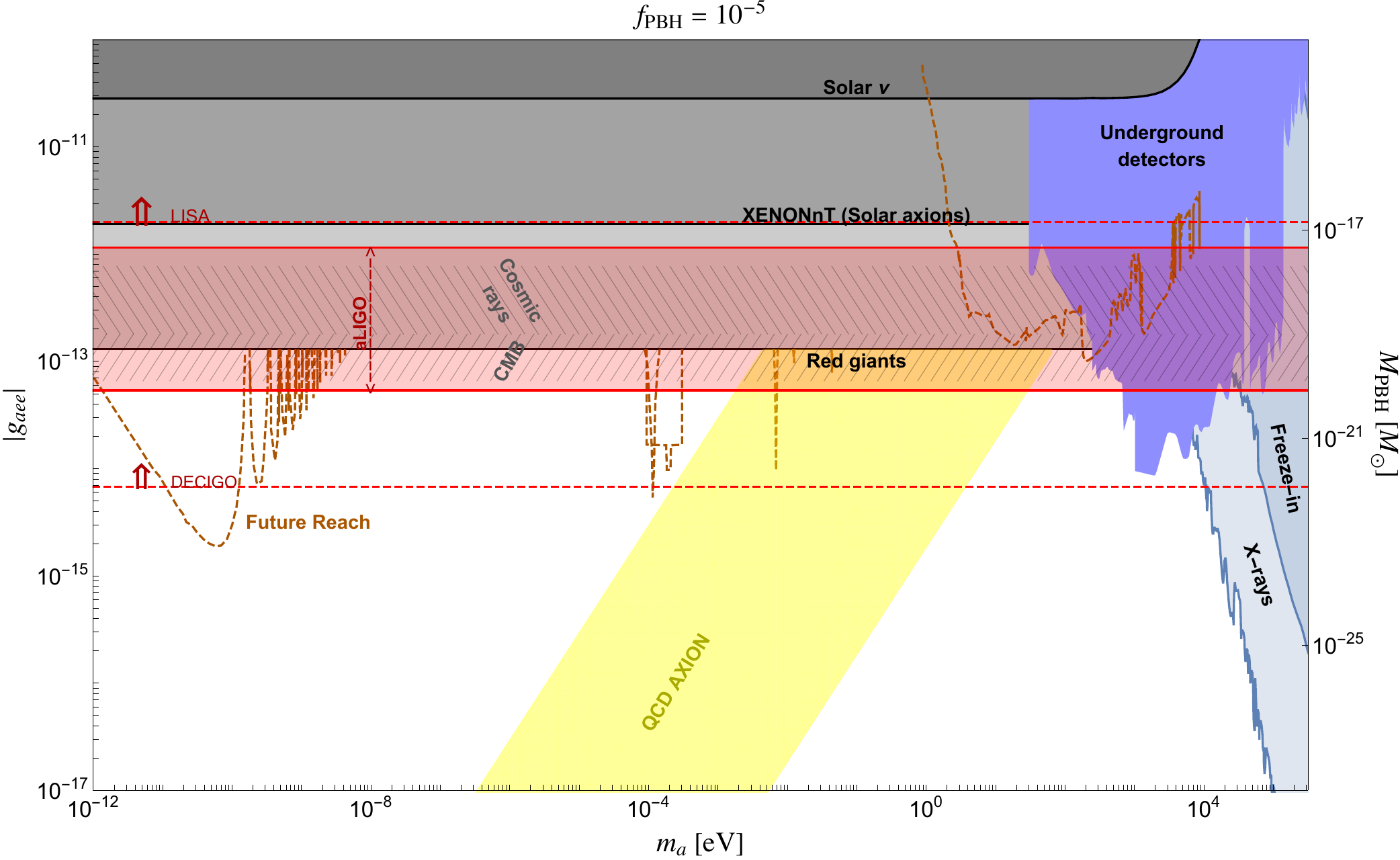}
%\vspace*{-0.5cm}
\caption{Same as Fig.~\ref{fig:axionelectron_fPBH1} but for $f_{\rm PBH} = 10^{-5}$. Only the PBH bounds become weaker, while all other bounds remain the same. } 
\label{fig:axionelectron_fPBHm5}
\end{figure}

\begin{figure}[h!]
\centering
\includegraphics[width=\textwidth]{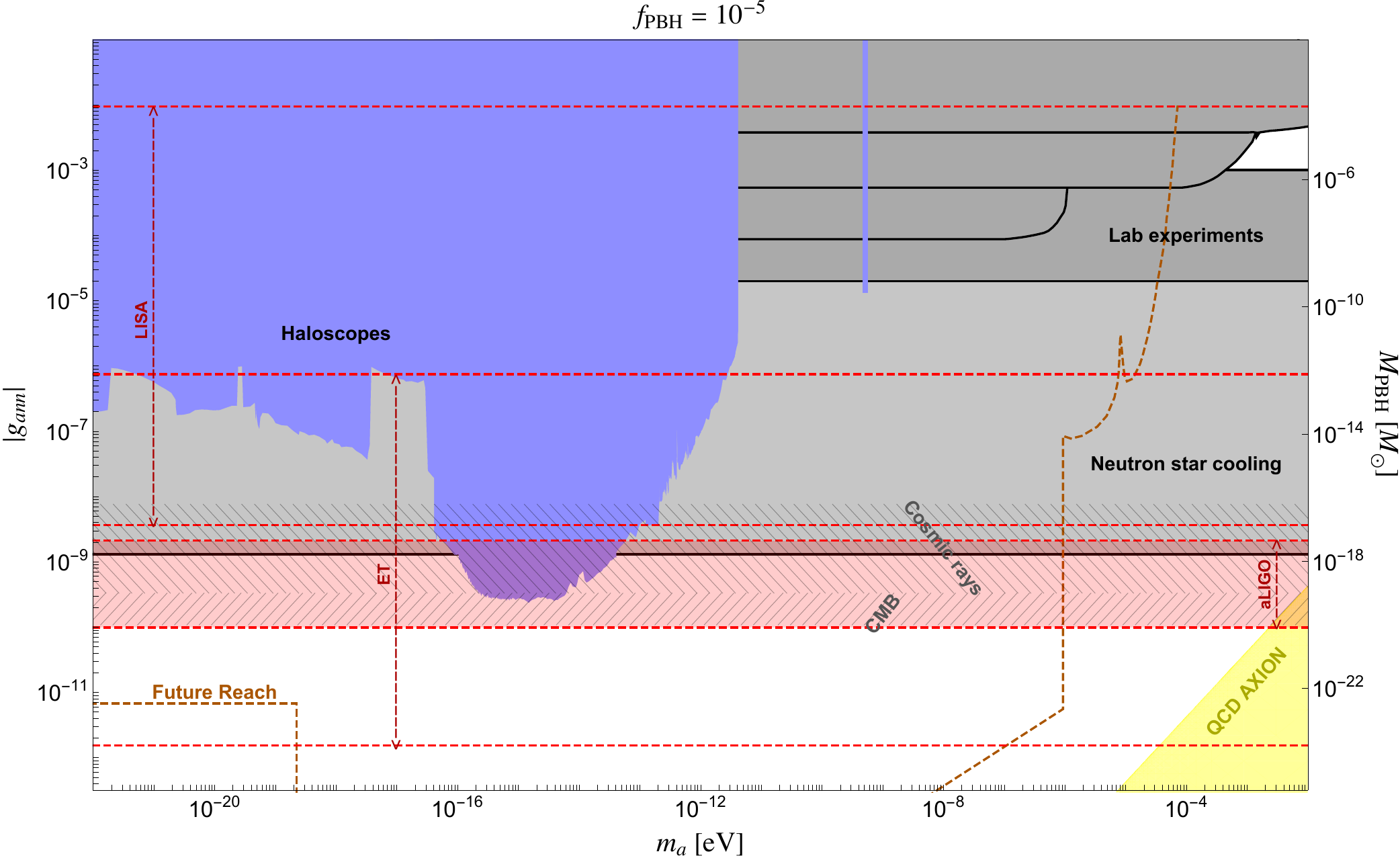}
%\vspace*{-0.5cm}
\caption{Same as Fig.~\ref{fig:axionneutron_fPBH1} but for $f_{\rm PBH} = 10^{-5}$. Only the PBH bounds become weaker, while all other bounds remain the same.} 
\label{fig:axionneutron_fPBHm5}
\end{figure}

In Figs.~\ref{fig:axionphoton_fPBHm5}, \ref{fig:axionelectron_fPBHm5} and \ref{fig:axionneutron_fPBHm5} we show the results for a smaller $f_{\rm PBH}=10^{-5}$. As expected, the PBH constraints become weaker, compared to the GW and other ALP constraints, which remain unchanged from Figs.~\ref{fig:axionphoton_fPBH1}, \ref{fig:axionelectron_fPBH1} and \ref{fig:axionneutron_fPBH1}, respectively. 


%%%%%%%%%%%%%%%%%%%%
\bibliography{Bibliography}
\bibliographystyle{JHEP}
%%%%%%%%%%%%%%%%%%

\end{document}